# The Theory and Applications of Parametric Excitation and Suppression of Oscillations in Continua: State of the Art

IVAN V. KAZACHKOV
[1]Department of Applied Mathematics and Informatics,
Nizhyn Gogol State University, UKRAINE
https://en.wikipedia.org/wiki/Nizhyn_Gogol_State_University
[2]Department of Energy Technology, Division of Heat and Power
Royal Institute of Technology, Brinellvägen, 68, Stockholm, 10044, SWEDEN
Ivan.Kazachkov@energy.kth.se    http://www.kth.se/itm/inst?l=en_UK

*Abstract:* - The results by development of physical, mathematical and numerical models for parametric excitation and suppression of oscillations on the interfaces separating continuous media, for carrying out computing, physical and natural experiments by revealing the new phenomena and parametric effects, and for their use in improvement the existing and creation the perspective highly efficient technological processes are presented. Scientific novelty of this work consists in development of the theory and applications of parametric excitation and suppression of oscillations on the boundaries of continua on the samples of three tasks' classes: flat and radial spreading film flows of viscous incompressible liquids, conductive as well as non-conductive ones; surfaces of phase transition from a liquid state into a solid one; and heterogeneous granular media. The external actions considered are: alternating electromagnetic, vibration, acoustic and thermal fields. Along with linear the non-linear parametric oscillations are investigated (including strongly non-linear) too and the results of theoretical studies are confirmed and supplemented with the corresponding experimental data. The general and specific peculiarities of parametrically excited oscillations and the new parametric effects revealed are discussed for technical and technological applications. First the general statement and substantiation of the problems studied is considered, and then the various parametric oscillations in continua are analyzed from common methodological base. Also the assessment of a current state of the problems, analysis of their features, prospects of further development and the main difficulties of the methodological, mathematical and applied character are presented.

## 1 Actuality of parametric oscillations

Actuality of the problem is caused by intensive development of the new technologies, creation of the devices and mechanisms of high efficiency and profitability based on the use of parametrically controlled oscillations in continua (working body), as well as based on stability of the system impossible in the absence of different effective suppression of oscillations.

The periodic or quasi-periodic influences having regular or casual character are widespread in technological installations and processes [1-7]. These are fluctuations of a field (electric, magnetic, acoustic, temperature, vibration, etc.) or such conditions under which the acyclic changes of certain parameters in a system cause oscillations of other parameters causing, in turn, parametric oscillations of the system.

Parametric oscillations in continua, especially on the boundaries separating different media (solid, liquid, gaseous), attracted the attention of many researchers due to their originality and widespread in the nature, technological and technical processes, etc. Significant development of the theory of parametric oscillations was launched in 19-th century by the fundamental works of Faraday, Savart, Magnus, Plateau, Rayleigh, Benard, Chernov, Cauchy, Korteweg, Reynolds, Ostrogradsky, and other classics of science.

Later on the theory was intensively developed by Bohr, Weber, Bussman, Haenlein, Taylor, Taylor, Chandrasekhar, Schlichting, Ohnesorge, Lin, Tollmien, Magarvey, Benjamin, Petrov, Landau, Kapitza, Zeldovich, Lavrentiev, Kutateladze etc.

The remarkable contribution to development of various areas of the theory of Eigen and parametric oscillations in continua and their use in engineering and technology was brought by Lighthill, Whitham, Joseph, Kramer, Ostroumov, Rakhmatulin, Nakoryakov, Nigmatullin, Butkovsky, Samoylenko, Gelfgat, Kolesnichenko, Entov, Nakorchevsky, Vesnitsky, Sukhorukov, Ostrovsky and others.

Parametric excitation and suppression of oscillations in continua gives promise on emergence of the new physical principles for creation of the processes or considerable improvement of possibilities and effectiveness of the existing and projected installations. The methods based on use of the strong (resonant) effects allowing developing the new energy- and resource-saving technologies important for intensive development of economy are especially perspective.

Nowadays many parametrical instability phenomena in the flows of gases and liquids complicated by phase transformations and chemical reactions [8-14] are well known in magneto-hydrodynamics (MHD) and physics of plasma [15-23], thermo- and hydrodynamics of the granular and underground natural systems [24-28], Biology [29-31] and others [32-35].

## 1.1 The research problem

Parametric oscillations in continua represent great theoretical interest and constantly extending practical applications, which, in turn, stimulate intensive development of the theory. Both for the theory, as well as for the practice, a case of parametric excitation of oscillations and a case of parametric suppression of casual or regular oscillations (stabilization of system, process) are equally important and have their own applications.

Possibility for sharp efficiency intensification of technological processes due to parametric oscillations in continua was revealed more than century ago: for example, Chernov (1879) for the first time noted positive influence of mechanical oscillations on a quality of crystallizing ingot [36]. However recommendations for a choice of parameters for external influences were often contradictory or absolutely absent. By control of crystallizing metal's structure the recommended range of fluctuations covers area of 1-250 Hz [2] and it is known that low-frequency mechanical, electromagnetic, thermal oscillations or their combinations most often crush the structure of metal making it smaller 2-10 times [24,37]. But the pulsing thermal field can lead to strengthening of orientation of the big size crystals.

Temporal parameters' modulation in continua influences stability of the physic-mechanical and chemical processes and can lead to stabilization of the certain kind instabilities [6,20-22,32,38,39] or, on the contrary, to excitement of instability [2,16,17,41-53]. Therefore the processes of parametric excitation and suppression of oscillations in continua, generally speaking, are closely interconnected, especially in case of strong non-linearity of the processes and in a presence of such complicating factors as heterogeneity of physical properties of media, relaxation, etc.

Intensive development of technology for generation of parametric oscillations in continua is caused by successful application in many areas: control of structure in MHD-flows [18,54,55], intensification of chemical and technological processes [56-58], localization of heating [59-63]. Considerable spreading was gained by vibration and acoustic methods in intensification of technological processes [1,5,9,41,64-69], which are also not completely studied in detail yet [58,71-73].

## 1.2 Classification of parametric oscillations

On a nature of impact on continua (technological processes), parametric oscillations may be classified as following: optimizing, intensifying, transforming and stabilizing. The first ones allow just improving the process or its some parts (acoustic granulation and centrifugation, etc. [5]). The second class increase a speed of process or its parts (the MHD-granulation [34,73], acoustic dissolution). Then the third class oscillations lead to receiving essentially new regimes or processes impossible in absence of parametric action: vibration dispersing of liquids, crushing of structure of an ingot, spouting of drops from a surface of the vibrating layer of liquid or of the liquid under acoustic field, etc. Finally the fourth class give a chance to carry out stable regime of process in continua (decrease in hydraulic resistance to moving bodies in a liquid by stabilization of the laminar mode of flow [74-76], ensuring running the chemical reactions with unstable parts, etc.). Obviously, a strict separation among these four types of parametric impact on the processes in continua doesn't exist.

The *effects* that are made by parametric oscillations in continua may be the ***first order*** (small-amplitude oscillations determined by frequency, intensity and speed of spreading in a medium) and the ***second order*** (powerful perturbations of medium causing the non-linear phenomena, violation of continuity, etc.).

Despite abundance of works on separate classes' parametrically excited (suppressed) oscillations in continua, research of a problem of parametric excitation and suppression of oscillations on the interfaces of continua is still in an initial state. Therefore the subject of researches mainly consists in the specified problem being considered on the example of specific classes of the tasks, e.g. three classes we studied were: film flows under prevailing contribution of the inertia forces, boundaries of the phase transition between liquid and solid states in a channel flow, granular gas-saturated media, with different types of parametric actions as for example in our works: electromagnetic and thermal fields, vibrations, etc. Parametric oscillations are considered mainly of the third (transforming) and the fourth (stabilizing) types (sometimes - the second type).

The main attention is paid here to revealing the opportunities of forecasting the reaction of continua to various external influences, especially – resonant ones. The parametric resonance oscillations can form a basis for creation the essentially new highly effective power- and resource-saving technologies, e.g. receiving powders and granules by means of the film MHD- and vibration type granulation machines [73,77,78]. The parametric oscillations having effect of both the first order, as well as the second order on the interfaces of continua are considered.

### 1.2.1 The specific of the problems considered

For the studied phenomena essential influence of regularities of spatial change of parameters of the continua owing to what parametric oscillations, generally speaking, are represented by the fields of various physical nature described by the partial differential equations (PDE) is characteristic. Therefore according to *the Law of Requisite Variety* by Ashby [79], the external influences have also to be the fields distributed in space and in time and described by the differential equations of mathematical physics.

External action can be transferred to each point of the continua (volumetric control) or to its boundary (boundary control). And the most effective impact on the processes in continua is hardly realized: so, the alternating electromagnetic field penetrates into conductive medium just on a distance of the skin-layer thickness, the acoustic wave also fades in a medium, and so on.

Comparably easily realized is boundary control in continua, the theory of which was developed by Butkovsky with colleagues [24,80], Sirazetdinov [81,82], Ladikov [21,22], and other researchers.

For electromagnetic control of the processes in continua Samoylenko proposed the linear layered and fibrous artificial media possessing high resolution (reaction of the medium to external action can be as much as close to the delta function) [83].

The theory of such systems was developed in [20,84]. On a nature of operating influences the problems of boundary control are subdivided into three classes [21] (control of the mass, impulse and energy fluxes), in each of which there are three types of control: with feedback, programming, and determination of dynamic properties of object by boundary values of its parameters.

This paper is mainly devoted to the problems of the second (disintegration of the film flows by parametric oscillations) and the third (stabilization of the boundaries of phase transition of thin films of a solid phase) classes with the first and the second types of influences. Thus in problems of parametric suppression of oscillations on the boundaries of phase transition the crucial importance belongs to an energy exchange while at the electromagnetic or vibration initiation of a surface's oscillations in film flows of viscous liquid the main role is played by an impulse exchange.

The open-loop and closed-loop control systems are applied to excitation and suppression of parametric oscillations in continua (with feedback): the first ones - when the return influence of the continua on the actuation device is negligible (for example, electromagnetic influence on medium in case of small magnetic Reynolds numbers), the second - for non-linear media and stabilization of the fast-proceeding processes, etc. Thus, at the first stage of researches in both cases it is necessary to reveal regularities of boundary impact on technological process, on the second stage - to define the structure and parameters of the device providing the demanded impact on continua.

In the modern continuum mechanics differing in merge to many areas of physics, penetration into various technical and biological systems with broad application of mathematical modeling and computers, the problems of the programmed open-loop control in touch with improvement of structure of ingots [37,38], vibration mixing and intensification of heat transfer [2,4], stabilization of technological processes [21,32,85], parametric wave excitation on liquid surfaces [49,86,87], stabilization of a flow in a boundary layer using suction and blow-in [57,76], and many others [35,88-93] have been considered. Many new tasks were solved during the recent decades [70,94-104].

### 1.2.2 Parametric action and problem formulation

In general the problem of parametric excitation and suppression of oscillations in continua can be formulated as follows. After schematization of the physical phenomenon, allocation of its most essential and minor parameters and creation of physical and mathematical models with the subsequent choice optimum from them (in a certain sense) it is necessary to investigate regularities of system perturbations' development in space-time.

It leads to establishment of the correlations of a type $F_n(A_j, \omega_j, \vec{k}_j; \vec{A}_j^*, \omega_*, \vec{k}_*) = 0$, where $A_j, \omega_j, \vec{k}_j$ are the amplitude, frequency and a wave vector of $j$-th perturbation, respectively, $\vec{A}_j^*$ is a vector of the perturbed parameters of the system (process) of dimension $J$. Asterisks noted parameters belong to Eigen fluctuations of a system, $F_n$ are the sought relations of parameters (e.g. differential), $n=\overline{1,N}$, $N \geq J$ (by $N<J$ there is an uncertainty of a task).

The behavior of continua is defined by parameters of Eigen fluctuations and external influences. Therefore the problem of excitation (suppression) of parametric oscillations can be reduced to determination of physically realized parameters $A_j, \omega_j, \vec{k}_j$ providing the necessary mode, or the set type of a vector function $\vec{A}_j^*(x, y, z, t)$.

In particular, in case of parametric oscillations on the interfaces separating different media it is often possible to reduce a task to solution of the equation for boundary oscillations (e.g. free surface of film flow, crystallization surface, etc.) and solution of the equation alternating external exciting force (field).

Problems of suppression and excitation of oscillations of the set type in continua demand control of all other perturbations as the energy pumping in a certain wave mode in non-linear systems can lead to its transfer into another modes, which may lead to a strongly complicated regime.

The *purpose* of our research was development of a set of physical, mathematical and numerical models for parametric excitation and suppression of oscillations on the interfaces separating continuous media, for carrying out computing, physical and natural experiments by revealing the new phenomena and parametric effects, and for their use in improvement the existing and creation the perspective efficient profitable technologies.

Scientific novelty of this work consists in development of the theory and applications of parametric excitation and suppression of oscillations on the boundaries of continua on the samples of three tasks' classes: flat and radial spreading film flows of viscous incompressible liquids, conductive as well as non-conductive ones; surfaces of phase transition from a liquid into a solid state; and heterogeneous granular media.

The external actions considered are: alternating electromagnetic field, vibration, acoustic and thermal fields. Along with linear ones also the non-linear parametric oscillations were investigated (including strongly non-linear) and the results of theoretical researches were confirmed and supplemented with the corresponding experimental data. The general and specific peculiarities of parametrically excited oscillations and the new parametric effects came to the light.

Based on the general statement of the problems studied and their substantiation made the various parametric oscillations in continua are considered from common methodological base. The assessment of a current state of the problems, analysis of their features, prospects of further development and the main difficulties of the methodological, mathematical and applied character are given.

### 1.2.3 Peculiarities of parametric oscillations in the film flows and crystallizing surfaces

The class of problems on parametric excitation and suppression of oscillations on the surfaces of the flat and radial film flows of conductive and non-conductive liquid, both free surface as well as restricted by some plates, carrying-out in non-conductive viscous incompressible liquid under impact of vibration, electromagnetic and acoustic fields in the form of progressive or standing waves, have been studied.

Then the researches of parametrically excited and suppressed oscillations on the boundaries of phase transition from a liquid state into a solid (generally in relation to stabilization of a surface of thin solid layer, which may be melted in some local regions, by means of electromagnetic fields and thermal regulators - automatic control systems) have been done as another class of tasks.

Parametric oscillations on the boundaries separating in a space phases of the heterogeneous system prone to temperature perturbations were considered with account of the non-linearity of the processes and physical properties of media, as well as with account of heterogeneous effects.

Experimental confirmation of the consistent patterns determined theoretically and also identification of the new parametric effects on physical models and natural facilities is presented in the final chapter.

### 1.3 The novelty of scientific field by parametric oscillations and their use

This *new scientific field* represents theoretical research on physical and mathematical models and experimental installations for the various parametric oscillations on the interfaces of continua and identification of the new effects and actions causing the essential technological applications.

The mathematical models have been built as much as possible simple but adequately reflecting if not all studied process, then at least its separate, most essential features. Complex non-linear models are realized by means of numerical methods on the computer. Models, generally rather simple, are convenient for computational experiments and possess high precision even on finite-difference numerical grids having rather small number of knots (some of them allow full realization or nearly full realization on personal computers).

*Validity and reliability* of the results obtained follow from validity and reliability of the initial theoretical base on parametric oscillations in continua and the models of continua applied, from strict substantiated statements of the tasks solved and the methods for their solution, experimental proof of the results and their comparison to the known limit cases.

*Practical value* of the work consists in the developed physical and mathematical models of parametric excitation and suppression of oscillations on the interfaces of continua. They allow calculating the concrete physical situations and choosing the technological solutions providing significant increase of quantitative and/or qualitative levels of the known devices or giving the chance for creating the *essentially new processes and devices*.

**1.3.1 New processes based on the results obtained**
The *new technological processes* based on the **discovered new phenomena** are provided here and some patents on the invented methods and devices for materials' granulation together with the results of their implementation into industry are described. In particular, description of the film granulator for magnesium and its alloys, which doesn't have analogs in world practice developed and introduced by the author together with employees is given.

All three classes of the tasks considered are important for this granulation machine: parametric control of liquid metal film disintegration (dispersing knot), stabilization of phase transition boundaries (protection of the channel walls against destruction and, at the same time, protection of the liquid metal in a channel against pollution with materials from the channels walls), thermo-hydrodynamics of the granular medium (selection of optimum cooling regime for granules and achievement of highest cooling rate for the drops by their solidification).

The processes studied have been described by partial differential equations (PDE) for mass, impulse and energy conservation, for homogeneous, as well as heterogeneous media. The equations for electromagnetic field and some known equations of the states were applied as well.

The mathematical methods applied were the theory of integral transformations, averaging the differential and integral-differential operators, reductive perturbation method for the non-linear PDE array, numerical methods for non-linear PDE (e.g. split method), physical modeling of the fields and media interaction, etc. Such combination of rather different analytical and numerical methods together with physical modeling and practical testing of the phenomena and devices allowed investigation the features of the new phenomena as much as possible in detail.

On an apt expression of the Nobel laureate on physics S. Chandrasekhar [6], "in magneto-hydrodynamics it is so easy to be mistaken that it isn't necessary to trust result of long and difficult mathematical calculations if it is impossible to understand it physical sense; at the same time it is impossible to rely on a long and difficult chain of physical arguments if it is impossible to show it mathematically". These words can be carried fully to a problem of parametric excitation and suppression of oscillations in continua investigated by us and implemented into practice.

**1.3.2 The concept of investigations made**
The concept accepted here was to investigate from united methodological positions a problem of parametric excitation and suppression of oscillations in three various systems (film flows, boundaries of phase transition and granular media), and to reveal both peculiar features for each of them and the general regularities too.

Both mathematical and physical modeling of the phenomena was applied to expand the areas of the studied phenomena and their comparative analysis in crossing areas, receiving data on adequacy of the constructed mathematical models and reliability of the technique applied. The results include:
1. Theory of parametric excitation and suppression of oscillations on the film flow surfaces under action of electromagnetic fields and vibrations.
2. Theoretically revealed and experimentally proved the new phenomena of soliton-like and shock-wave decay of the film flows into drops.
3. Theory of parametric control of oscillations on the boundaries of phase transition from melted to solid state on the channel walls.
4. The results for thermal hydraulic oscillations in granular saturated media: phenomenon of dissipative processes' localization and local

abnormal heating due to non-linear heat conductivity of vapor flow, etc.
5. Experimental studies and natural experiments with the processes (and facilities on their base) on parametric control for film flow waves and the new created devices and processes.

The results obtained are new in this field both in a theory and practical applications. Therefore our decision to publish these results came with a hope to present some prospective achievements, which may be of interest for scientists and industrial engineers.

## 2 Scientific and engineering problems on parametric oscillations

Intensive development of the modern industrial technologies and equipment demands for research of parametrically excited (caused by the periodic exciting force) in continua. The examples are as flows: dynamic stability of elastic systems, fluctuations of plates, covers, rods, oscillation of liquids in the vibrating vessels, in pipelines, pumps, etc. Nowadays numerous ways of parametric excitation and suppression of oscillations in continua, in particular, in the liquids are known. For example, physics of the superficial phenomena and thin films [10,33,42,85,89,90,105-121].

The new directions in modern natural sciences [35,50-52,67,86,97,122-125] appeared in touch with research of the non-linear processes by different physic-mechanical and chemical nature: thermo-hydrodynamic and magneto-hydrodynamic instability, catastrophes, bifurcations, self-oscillations, etc. In recent decades it was established that an ideal way of energy transfer in non-linear systems is a soliton [14,126-131], which mathematical description for shallow water was given by Korteveg-De-Vriz [131]. Lavrentyev presented the strict proof of their existence for a liquid of finite depth [132].

Steady solitary waves are formed as a result of mutual compensation of the non-linearity and dispersion and combine wonderfully properties of a particle and a wave [29,31].

Three new phenomena on parametric wave excitation have been revealed and studied by us both theoretically and experimentally [46-49,73,116, 132-136], and then implemented into industrial practice of the new materials creation based on the high speed cooling of the small drops, which produce unique properties of the metals obtained of it [77,78,137-141].

### 2.1 Phenomena of thermal waves

A number of the new interesting phenomena were found when studying thermal waves [66,143,144]. Here the non-linearity caused by dependence of physical properties of the medium on coordinates and time, in particular, heat conductivity coefficient dependent on temperature, which comes to zero at the front a thermal wave. Presence of some sources and severs in a medium leads to inertial delay of the speed of perturbation spreading.

The revealed mechanism of volumetric heat absorption [59] was used as one of possible explanations for thermal self-isolation of a fireball [144]. Mathematical models of non-linear heat conductivity with adaptation on concrete cases allow describing a wide class of diffusion problems, distribution of thermal and electromagnetic fields.

Some features of the considered classes of tasks have been already noted above, the applied mathematical models of continua and the ways of excitation and suppression of oscillations in continua are described and analyzed here (the main attention is paid to the boundary control in continua). Possible difficulties of methodological, mathematical and applied character are in a focus too. Besides, the statement of a number of unsolved tasks making a subject of our researches (present and partly the future ones) is given as well.

### 2.2 Development of the physical and mathematical models of continua

Creation of physical and mathematical models of continua and subsequent research of the physic-mechanical and other processes running in them is most often based on the hypotheses of a continuity of medium and continuous *n*-times differentiability of all functions describing parameters of continua, almost everywhere except for separate points, lines or surfaces on which gaps are allowed. It allows using phenomenological approach applying the developed methods of the mathematical analysis and mathematical physics.

However, though phenomenological approach allowed solving a set of the problems in continuum mechanics, which became already classical and is nowadays one of the most often applied, it is necessary to account that many processes in continua don't satisfy this physical model.

For example, in turbulent flow the velocity acceleration fields aren't described in a class of continuous or nearly everywhere continuous functions, and the film flow compressed in the direction of a tangent to a free surface becomes nowhere differentiable in this direction even if remains continuous (a surface saw-tooth with teeth, perpendicular to it). The other examples are spraying and cavitation, for which individualization of volumes is impossible since the continua turns into a set of the free points.

In heterogeneous media the fields of velocities, temperatures, etc. are fractured and combination of two various fields in one continuum, a polysemy of parameters belong not to the individualized point of a medium but to a space point in which the individualized points of parameters are combined in.

Statistical approach and various options of variation methods [146-150] are applied to non-classical problems of the continuum mechanics, which aren't satisfying to hypotheses of the phenomenological theory. In view of mathematical complexity statistical (microscopic) approach is often used for justification of phenomenological (macroscopic) models of continua if only it isn't unique: the discharged gases, plasma, etc. Strictly speaking, in the nature there are no real continua, however the continuity hypothesis describes them well at the macro level and the model of continua allows using the powerful theory of calculus for continuous functions.

For the systems, which do not answer continuity hypothesis of an occupied space, the fractal [151] theory (objects of fractional dimension) and the developed integral and differential calculus of arbitrary order [152,153] (not only integer derivations and integrals as in classical calculus) are more adequate. Here we generally use phenomenological approach at which creation of mathematical models of continua is based on an assumption that each point of the medium (physically infinitesimal volume), which physical and mathematical state is characterized by a set of the defining parameters introduced on the basis of experimental data and theoretical investigations.

Now from a position of phenomenological approach such set of mathematical models for various tasks' classes accounting their specific features has been developed that a need ripened for their systematization and development the basic principles of mathematical modeling of processes in continua. Therefore when developing the new complexes of mathematical and numerical models it is necessary to proceed from the modular principle allowing to unify as much as possible a process of modeling and to facilitate the use of mathematical and computer numerical models by various researchers in various tasks.

### 2.2.1 The equations of dynamics of continua

According to phenomenological approach, creation of mathematical models of continua is based on an assumption that each point of the medium (physically infinitesimal volume), which physic-mechanical state is characterized by a set of the defining parameters introduced on the basis of experimental and theoretical data or statistically averaged functions (temperature, for example).

The general equations of dynamics of continua by any structure (including the heterogeneous mix considered without phase interaction, which can be not taken into account when studying the movement of heterogeneous system as a uniform complex continuous medium), may be represented in a form:

$$\partial \rho / \partial t = -div(\rho \vec{v}), \quad (1)$$

$$\rho(\partial \vec{v} / \partial t + \vec{v}\nabla \vec{v}) = divP + \rho \vec{F} - \sum_{j=1}^{N} \vec{v}\nabla(\rho_j \vec{v}_j), \quad (2)$$

$$\rho(\partial e / \partial t + \vec{v}\nabla e) = div(\vec{q} + P\vec{v}) + \rho \vec{F}\vec{v} + \quad (3)$$
$$+ \sum_{j=1}^{N}\left[\rho_j \vec{F}_j \vec{v} - div(\rho_j e_j \vec{v})\right],$$

where $\rho$ - density of heterogeneous medium, $\vec{v}$ - velocity vector for heterogeneous medium, $t$ - time, $P$ - stress tensor, $\vec{F}$ - volumetric force, $\rho_j, \vec{v}_j$ - parameters of the medium's components (similar for the other parameters), $e$ - specific density of energy, $q$ - specific volumetric energy influx.

The first equation (1) is mass conservation law, the second (2) - balance of an impulse, the third (3) – energy conservation. For reversible processes the uncompensated warmth is equal to zero. Except internal energy and entropy the other functions of state, as well as some additional thermodynamic relations are used.

### 2.2.2 The exchange of mass, impulse and energy between phases of heterogeneous medium

In case of heterogeneous medium when the mass, impulse and energy exchange between phases inside the volume or at the boundaries must be taken into account, the terms on such exchanges between phases of heterogeneous mix in the equation array (1)-(3) must be explicitly specified. Namely this makes the main problem in mechanics of heterogeneous media since it is in the most cases unknown. One needs to clarify the intensities of the mass, impulse and energy exchange between the phases of heterogeneous continua.

Each phase occupies some part of each elementary volume of the heterogeneous medium: the volumetric contents of $N$ phases $\alpha_j$ satisfies the equation $\sum_{j=1}^{N}\alpha_j = 1$, density of the medium is expressed through the real densities of phases by a formula $\rho = \sum_{j=1}^{N} \rho_j \alpha_j$. At each point of the heterogeneous medium $N$ parameters are defined relating to the continuum (densities, velocities, temperatures, etc.). Set of continuums, each of which corresponds to its phase and fills the same volume, is called the multi-speed continuum.

Summation of the equations by all phases of the mix gives the equations of the heterogeneous medium taken as a uniform system, without account the internal structure. Such model doesn't show features of interfacial interaction in heterogeneous mix. Contrary to it, accounting an interaction of the phases' macroscopic inclusions results in need for account the conditions of joint deformation and movement of phases, influences of a form and the amount of inclusions, their distributions in space, phase transformations, etc.

If physic-mechanical processes in the continua are rather precisely described by continuous or nearly everywhere continuous functions of coordinates and time, it is possible to replace the system of integral conservation equations to the corresponding differential equations. However for the real continua prone to the external influences, the classical methods can be unacceptable owing to what the variation and numerical methods based on use of the integral correlations are remaining useful in case of fractured fields and media if integration by Riemann is replaced by integration by Lebesgue.

### 2.2.3 The deformation of heterogeneous medium and interfacial interactions in a medium

In a region of continuous or nearly everywhere continuous movement of a continuum, using Gauss-Ostrogradsky's formula, it is possible to pass from the integral balance equations to the differential equation array describing thermo-hydrodynamic processes in the heterogeneous medium taking into account the joint movement of phases and the interfacial mass, momentum and energy exchange. The main obstacle in use this system by mathematical modeling of heterogeneous media is caused by need of a specification of laws of phases' interactions that is the extremely difficult.

The law of deformation of the heterogeneous medium depends not only on velocity fields, pressure, and temperature of phases; therefore determination of regularities of interfacial interaction even for special cases is very complex challenge. And still the accounting of fields' ruptures on the boundary interfaces is absolutely necessary for some practically important tasks.

At rather weak manifestation of interfacial interaction in the heterogeneous media, for the description of the processes happening in it, the system of differential equations obtained from the conservation equations by each component of the heterogeneous mix through the summation on all mix may be used.

### 2.2.4 The balance equations

The balance equations of an impulse and energy depend remarkably on relative movement of phases inside heterogeneous mix. Here later on from the mass forces only gravitational and electromagnetic ones are mainly considered. And inflow of the external energy (the so called Joule heat) or vibration energy of action transmitted through the boundary interface is accounted. Therefore differential equation array must also include the field equations:

$$div\vec{B} = 0, \quad \partial \vec{B}/\partial t = -rot\vec{E}, \quad \vec{j} = \partial \vec{D}/\partial t + rot\vec{H}; \quad (4)$$

$$\rho d\vec{v}/dt = divP + \rho\vec{g} + \rho_e\vec{E} + \vec{j} \times \vec{B}, \quad \rho_e = div\vec{D}; \quad (5)$$

$$\rho de/dt = \vec{E}' \cdot \vec{j}' + \rho\vec{g}\vec{v} + \Phi + div(\vec{q}^{вш} - \vec{q}^{вн}) - pdiv\vec{v}, \quad (6)$$

where $\vec{E}' \cdot \vec{j}'$ is the Joule heat, $\Phi$ is the dissipation function, $\vec{E}' = \vec{E} + \vec{v} \times \vec{B}$, $\vec{j}' = \vec{j} - \rho_e\vec{v}$ - vectors of electric field and current density in a coordinate system connected to the considered volume of the moving continua.

The stress tensor for the Newtonian fluid (in the paper only such ones are mainly considered) is presented in the form of the sum of a spherical tensor and a deviator: $P_{ik} = -p\delta_{ik} + \tau_{ik}$ where by a "mute" index supposed to be summation, $\delta_{ik}$ - the Kronecker symbol. Dissipative function $\Phi$ has a form $\Phi = \tau_{ik}v_{i,k}$, where $i,k$ are the so-called "mute" indexes, $v_{i,k}$ - derivative from $i$-th velocity component by the $k$-th coordinate. The heat flux is taken by the Fourier law: $\vec{q} = -\kappa\nabla T$, $\kappa$ - heat conductivity coefficient.

### 2.2.5 The closing relations and laws

The Ohm law for density of the electric current is:

$$\vec{j} = \rho_e\vec{v} + \gamma_e(\vec{E} + \vec{v} \times \vec{B}). \quad (7)$$

The system (1)-(3) can be transformed to a divergent form that is very important for numerical simulation. For its closing, except the above-stated, there are used also some other defining correlations, as well as the empirical dependences of physical characteristics of media from the parameters of state (pressure, temperature), etc.

For instance, $\kappa$ is, generally speaking, function of pressure and temperature, and for linear materials without polarization the constitutive relations are connecting the vectors of magnetic and electric induction with intensity vectors of electric and magnetic fields through the equations:

$$\vec{B} = \mu_m \vec{H}, \quad \vec{D} = \mu_e \vec{E}. \qquad (8)$$

It is assumed that all continua considered here (homogeneous as well as heterogeneous) are two-parametric [155], so that their thermodynamic functions $e, p, s$ are determined by two parameters of state. This allow using the Gibbs correlation

$$de = Tds - pdV, \qquad (9)$$

where entropy for incompressible fluid is

$$s = c_V \ln \frac{p}{\rho^{c_{p/V}}} + const, \qquad (10)$$

therefore $de = c_V dT + const$, and for ideal gas with account of the Clausius-Clapeyron relation yields:

$$p = \rho RT, \quad c_p dT = c_V dT + d(p/\rho), \qquad (11)$$

Here $c_p$ is, in general, function of temperature.

By consideration of heterogeneous media it is supposed that properties of each phase are defined from a condition of filling with this phase the total volume. Temperatures of phases are introduced on the basis of a hypothesis of local thermodynamic balance within a phase. Further the non-stationary differential equations of non-isothermal movement of the viscous incompressible liquid (1)-( 3), (4)-( 6) (which is carrying out or non-conducting) in the Cartesian and cylindrical coordinate systems are used. Therefore it is expedient to write them here for rather general case at constants $\mu_m, \nu_m$ and variables $\rho, \mu, \kappa$. And in each considered case the ratios (7)-(11) concretizing a model of the continua are used.

## 2.3 The equations in Cartesian coordinates

Without limiting a generality, it is possible to consider that the axis is directed against gravity. Then, believing magnetic force prevailing the electric one, taking into account the above-mentioned, the system of differential equations (1.1.5), (1.1.6) can be written as:

$$\rho\left(\frac{\partial u}{\partial t} + u\frac{\partial u}{\partial x} + v\frac{\partial u}{\partial y} + w\frac{\partial u}{\partial z}\right) + \frac{\partial}{\partial x}\left(p + 0{,}5\mu_m |\vec{H}|^2\right) =$$

$$= \frac{\partial}{\partial x}(\mu \frac{\partial u}{\partial x}) + \frac{\partial}{\partial y}(\mu \frac{\partial u}{\partial y}) + \frac{\partial}{\partial z}(\mu \frac{\partial u}{\partial z}) +$$

$$\mu_m\left(H_x \frac{\partial H_x}{\partial x} + H_y \frac{\partial H_x}{\partial y} + H_z \frac{\partial H_x}{\partial z}\right),$$

$$\rho\left(\frac{\partial v}{\partial t} + u\frac{\partial v}{\partial x} + v\frac{\partial v}{\partial y} + w\frac{\partial v}{\partial z}\right) + \frac{\partial}{\partial y}\left(p + \frac{\mu_m}{2}|\vec{H}|^2\right) = \mu_m(H_x \frac{\partial H_y}{\partial x} +$$

$$+ H_y \frac{\partial H_y}{\partial y} + H_z \frac{\partial H_y}{\partial z}) + \frac{\partial}{\partial x}(\mu \frac{\partial v}{\partial x}) + \frac{\partial}{\partial y}(\mu \frac{\partial v}{\partial y}) + \frac{\partial}{\partial z}(\mu \frac{\partial v}{\partial z}),$$

$$\rho\left(\frac{\partial w}{\partial t} + u\frac{\partial w}{\partial x} + v\frac{\partial w}{\partial y} + w\frac{\partial w}{\partial z}\right) + \frac{\partial}{\partial z}\left(p + \frac{\mu_m}{2}|\vec{H}|^2\right) = \mu_m(H_x \frac{\partial H_z}{\partial x} +$$

$$+ H_y \frac{\partial H_z}{\partial y} + H_z \frac{\partial H_z}{\partial z}) + \frac{\partial}{\partial x}(\mu \frac{\partial w}{\partial x}) + \frac{\partial}{\partial y}(\mu \frac{\partial w}{\partial y}) + \frac{\partial}{\partial z}(\mu \frac{\partial w}{\partial z}),$$

$$\frac{\partial \rho}{\partial t} + \frac{\partial(\rho u)}{\partial x} + \frac{\partial(\rho v)}{\partial y} + \frac{\partial(\rho w)}{\partial z} = 0, \qquad (12)$$

$$\rho c_v\left(\frac{\partial T}{\partial t} + u\frac{\partial T}{\partial x} + v\frac{\partial T}{\partial y} + w\frac{\partial T}{\partial z}\right) + p\left(\frac{\partial u}{\partial x} + \frac{\partial v}{\partial y} + \frac{\partial w}{\partial z}\right) =$$

$$= \frac{\partial}{\partial x}(\kappa \frac{\partial T}{\partial x}) + \frac{\partial}{\partial y}(\kappa \frac{\partial T}{\partial y}) + \frac{\partial}{\partial z}(\kappa \frac{\partial T}{\partial z}) + \frac{\partial q_x^{вш}}{\partial x} + \frac{\partial q_y^{вш}}{\partial y} + \frac{\partial q_z^{вш}}{\partial z} +$$

$$+ \nu_m\left[\left(\frac{\partial H_z}{\partial y} - \frac{\partial H_y}{\partial z}\right)^2 + \left(\frac{\partial H_x}{\partial z} - \frac{\partial H_z}{\partial x}\right)^2 + \left(\frac{\partial H_y}{\partial x} - \frac{\partial H_x}{\partial y}\right)^2\right] +$$

$$+ 0{,}5\mu\left[\left(\frac{\partial u}{\partial y} + \frac{\partial v}{\partial x}\right)^2 + \left(\frac{\partial w}{\partial x} + \frac{\partial u}{\partial z}\right)^2 + \left(\frac{\partial v}{\partial z} + \frac{\partial w}{\partial y}\right)^2\right],$$

$$\frac{\partial H_x}{\partial t} = \nu_m\left(\frac{\partial^2 H_x}{\partial x^2} + \frac{\partial^2 H_x}{\partial y^2} + \frac{\partial^2 H_x}{\partial z^2}\right) + \frac{\partial}{\partial y}(uH_y - vH_x) +$$

$$+ \frac{\partial}{\partial z}(uH_z - wH_x), \quad \frac{\partial H_y}{\partial t} = \frac{\partial}{\partial z}\left(vH_z - wH_y\right) +$$

$$+ \frac{\partial}{\partial x}(vH_x - uH_y) + \nu_m\left(\frac{\partial^2 H_y}{\partial x^2} + \frac{\partial^2 H_y}{\partial y^2} + \frac{\partial^2 H_y}{\partial z^2}\right),$$

$$\frac{\partial H_z}{\partial t} = \frac{\partial}{\partial x}\left(wH_x - uH_z\right) + \frac{\partial}{\partial y}\left(wH_y - vH_z\right) +$$

$$\nu_m\left(\frac{\partial^2 H_z}{\partial x^2} + \frac{\partial^2 H_z}{\partial y^2} + \frac{\partial^2 H_z}{\partial z^2}\right), \quad \frac{\partial H_x}{\partial x} + \frac{\partial H_y}{\partial y} + \frac{\partial H_z}{\partial z} = 0.$$

The potential part of electromagnetic field (magnetic pressure) was separated in the equation array (12), therefore it is called the symmetric form.

## 2.4 Equation array in cylindrical coordinates

In a cylindrical coordinate system $(r, \varphi, z)$, the coordinate surfaces are cylinders $r = const$, semi-planes $\varphi = const$ and planes $z = const$, therefore differential equation array (12) transforms to:

$$\frac{\partial \rho}{\partial t}+\frac{\partial(\rho u)}{\partial r}+\frac{\rho u}{r}+\frac{\partial(\rho v)}{r\partial \varphi}+\frac{\partial(\rho w)}{\partial z}=0,$$

$$\rho\left(\frac{\partial u}{\partial t}+u\frac{\partial u}{\partial r}+\frac{v\partial u}{r\partial \varphi}+w\frac{\partial u}{\partial z}-\frac{v^2}{r}\right)+\frac{\partial}{\partial r}\left(p+\frac{\mu_m|\vec{H}|^2}{2}\right)=$$

$$=\mu_m(H_r\frac{\partial H_r}{\partial r}+\frac{H_\varphi \partial H_r}{r\partial \varphi}+H_z\frac{\partial H_r}{\partial z}-\frac{H_\varphi^2}{r})+\frac{\partial}{r^2\partial \varphi}(\mu\frac{\partial u}{\partial \varphi})+$$

$$+\frac{\partial}{\partial r}(\mu\frac{\partial u}{\partial r})+\frac{\partial}{\partial z}(\mu\frac{\partial u}{\partial z})+\frac{\partial(\mu u)}{r\partial r}-\frac{\mu}{r^2}\left(2\frac{\partial v}{\partial \varphi}+u\right),$$

$$\rho\left(\frac{\partial v}{\partial t}+u\frac{\partial v}{\partial r}+\frac{v\partial v}{r\partial \varphi}+\frac{w\partial v}{\partial z}+\frac{uv}{r}\right)+\frac{\partial}{r\partial \varphi}\left(p+\frac{\mu_m}{2}|\vec{H}|^2\right)=$$

$$=\mu_m\left(H_r\frac{\partial H_\varphi}{\partial r}+\frac{H_\varphi \partial H_\varphi}{r\partial \varphi}+H_z\frac{\partial H_\varphi}{\partial z}+\frac{H_r H_\varphi}{r}\right)+\frac{\partial}{\partial r}(\mu\frac{\partial v}{\partial r})+$$

$$+\frac{\partial}{r^2\partial \varphi}(\mu\frac{\partial}{\partial \varphi})+\frac{\partial}{\partial z}(\mu\frac{\partial v}{\partial z})+\frac{\partial(\mu v)}{r\partial r}+\frac{\mu}{r^2}(2\frac{\partial u}{\partial \varphi}-v),$$

$$\rho\left(\frac{\partial w}{\partial t}+u\frac{\partial w}{\partial r}+\frac{v\partial w}{r\partial \varphi}+w\frac{\partial w}{\partial z}\right)+\frac{\partial}{\partial z}\left(p+\frac{\mu_m}{r^2}|\vec{H}|^2\right)+\rho g=$$

$$=\mu_m\left(H_r\frac{\partial H_z}{\partial r}+\frac{H_\varphi \partial H_z}{r\partial \varphi}+H_z\frac{\partial H_z}{\partial z}\right)+\frac{\partial}{\partial r}(\mu\frac{\partial w}{\partial r})+$$

$$+\frac{\partial}{r^2\partial \varphi}(\mu\frac{\partial w}{\partial \varphi})+\frac{\partial}{\partial z}(\mu\frac{\partial w}{\partial z})+\frac{\partial(\mu w)}{r\partial r}, \quad (13)$$

$$\rho c_V\left(\frac{\partial T}{\partial t}+u\frac{\partial T}{\partial r}+\frac{v\partial T}{r\partial \varphi}+\frac{w\partial T}{\partial z}\right)+p\left(\frac{\partial u}{\partial r}+\frac{u}{r}+\frac{\partial v}{r\partial \varphi}+\frac{\partial w}{\partial z}\right)=$$

$$=\frac{\partial}{\partial r}(\kappa\frac{\partial T}{\partial r})+\frac{\partial}{r^2\partial \varphi}(\kappa\frac{\partial T}{\partial \varphi})+\frac{\partial}{\partial z}(\kappa\frac{\partial T}{\partial z})+\frac{\partial(\kappa T)}{r\partial r}+$$

$$+v_m\left[\left(\frac{\partial H_z}{r\partial \varphi}-\frac{\partial H_\varphi}{\partial z}\right)^2+\left(\frac{\partial H_r}{\partial z}-\frac{\partial H_z}{\partial r}\right)^2+\left(\frac{\partial H_\varphi}{\partial r}+\frac{H_\varphi}{r}-\frac{\partial H_r}{r\partial \varphi}\right)^2\right]+$$

$$+\frac{\mu}{2}\left[\left(\frac{\partial u}{r\partial \varphi}-\frac{\partial v}{\partial r}-\frac{v}{r}\right)^2+\left(\frac{\partial v}{\partial z}+\frac{\partial w}{r\partial \varphi}\right)^2+\left(\frac{\partial w}{\partial r}+\frac{\partial u}{\partial z}\right)^2\right]+$$

$$+\frac{\partial q_r^{ex}}{\partial r}+\frac{q_r^{ex}}{r}+\frac{\partial q_\varphi^{ex}}{r\partial \varphi}+\frac{\partial q_z^{ex}}{\partial z}=0,$$

$$\frac{\partial H_r}{\partial t}=\frac{1}{r}\left[\frac{\partial}{\partial \varphi}(uH_\varphi-vH_r)+r\frac{\partial}{\partial z}(uH_z-wH_r)\right]+$$

$$+v_m\left(\frac{\partial^2 H_r}{\partial r^2}+\frac{\partial^2 H_r}{r^2\partial \varphi^2}+\frac{\partial^2 H_r}{\partial z^2}+\frac{\partial H_r}{r\partial r}-\frac{2\partial H_\varphi}{r^2\partial \varphi}-\frac{H_r}{r^2}\right),$$

$$\frac{\partial H_\varphi}{\partial t}=\frac{\partial}{\partial r}\left(vH_r-uH_\varphi\right)+\frac{\partial}{\partial z}(vH_z-wH_\varphi)+$$

$$+v_m\left(\frac{\partial^2 H_\varphi}{\partial r^2}+\frac{\partial^2 H_\varphi}{r^2\partial \varphi^2}+\frac{\partial^2 H_\varphi}{\partial z^2}+\frac{\partial H_\varphi}{r\partial r}+\frac{2\partial H_r}{r^2\partial \varphi}-\frac{H_\varphi}{r^2}\right),$$

$$\frac{\partial H_z}{\partial t}=\frac{1}{r}\left[\frac{\partial}{\partial r}r(wH_r-uH_z)+\frac{\partial}{\partial \varphi}(wH_\varphi-vH_z)\right]+$$

$$+v_m\left(\frac{\partial^2 H_z}{\partial r^2}+\frac{\partial^2 H_z}{r^2\partial \varphi^2}+\frac{\partial^2 H_z}{\partial z^2}+\frac{\partial H_z}{r\partial r}\right),$$

$$\frac{\partial H_r}{\partial r}+\frac{H_r}{r}+\frac{\partial H_\varphi}{r\partial \varphi}=-\frac{\partial H_z}{\partial z}.$$

In the represented general form the partial differential equation (PDE) array (12), (13) has mostly heuristic meaning and is normally used for numerical simulation. In many cases these PDEs are rather simplified based on the features of the problem stated. But as concern to analytical solution such PDE array is normally substantially simplified using specific character of the problem stated and estimation of the terms in the equations.

Nowadays thanks to the remarkable progress of computers and numerical methods such PDE may be solved for many important tasks in a general statement. The numerical solution of many tasks in a general statement is still difficult even with the most powerful computers because the existing mathematical theory of numerical solution for the non-linear PDE is still inadequate.

Lack of strict estimates of the errors of numerical solutions for PDE systems, proofs of stability and convergence of numerical algorithms causes the necessity of the integrated approach to a problems' solution including the following main stages:

- strict mathematical research of the simplified linearized tasks (with attraction of heuristic substantiation and physical intuition),
- computational experiment on the computer,
- establishment of adequacy of physical and mathematical models to a real physical process.

Considerable difficulties in mathematical modeling of physic-mechanical processes in continua are connected with incompleteness of the theory of existence and uniqueness of PDE solutions describing the thermo- and magneto-hydrodynamic phenomena. And the theory is far from completeness as regards the equations, as well as their finite-difference and finite-element analogues.

Mathematical problems of convergence and stability of numerical schemes are well developed only for linear tasks while non-linear ones are still in many cases insufficiently studied. As the initial approximations the linear theories are used.

It is known that in many cases the solutions of the non-linear tasks strongly depend on accuracy and adequacy of statement of the boundary conditions. The destabilizing influence of boundary conditions at the solution of non-linear tasks makes the main significance owing to what the derivation of boundary conditions from physical laws and investigation of their influence on a character of the solutions obtained requires paying the attention.

## 2.5 The boundary-value problems

Statement of the boundary-value problems for parametric oscillations in continua requires specifying the type of parametrical influence: external electromagnetic field, vibration or thermal action, etc. Vibration can be characterized by a vector of vibration acceleration and the process can be considered in a moving coordinate system connected to the vibrating surface, which on a liquid spreads, considering an action of vibrations as oscillating body force.

Physical substantiation of the initial and boundary conditions is also important stage of researches. Generally it is necessary to set spatial distribution of the main parameters of medium at the initial moment and the corresponding conditions on borders of the considered space area. But in the problems of stability and wave spreading in a linear statement it is possible to seek solution in the form of superposition of progressive or standing waves. Then the initial conditions don't essentially influence a nature of process as far as the main is a question about regularities of waves' increase (decrease) but not absolute value of amplitude at the initial moment.

### 2.5.1 Linear case

In a linear theory the progressive waves

$$\tilde{q}_j = Q_j(y, z, \omega) \exp i(kx - \omega t), \qquad (14)$$

are spreading along the axis $x$ with velocity $\omega/k$. Here $Q_j$ is complex amplitude. Separation of the real part may be done in a final result.

### 2.5.2 Substantiation of the boundary conditions

Derivation and substantiation of the boundary conditions is definitely an important task to which it is given more and more attention [156] in connection with the problems of obtaining the missing conditions, and also appearing the singularities and influence of lack of boundary conditions on a nature of modeled processes.

It was established that adequacy of any boundary condition depends on other conditions (boundary and initial) and the accepted mathematical model, and it makes an essential feature of the computational experiment in each specific case. Thus, large number of the defining factors often complicates and sometimes does impossible an analytical research of parametric oscillations in continua or limits its application [101,156,157].

As shown by computational experiments, the question of adequacy of boundary conditions is connected to the type of difference schemes, ratio of the defining dimensionless criteria, etc. In analytical researches an adequacy of boundary conditions influences adequacy of the solution obtained. In computational experiment the boundary conditions define not only the accuracy of the solution obtained, but also stability of the computational algorithm, i.e. possibility of obtaining the solution.

Richardson gave (1910) accurate characteristic of a problem of adequate statement of the boundary conditions [158], but due attention to it began to pay only recently [159-165]. In most cases research of boundary conditions was carried out on computers with use the simple two-layer explicit finite-difference schemes for the boundary conditions of the first and the second type. Tasks with the mixed and non-linear boundary conditions present specific class [166], which is not studied seriously yet.

Influence of the boundary conditions on solution of the boundary-value tasks about parametric oscillations in continua is especially important in a presence of the complex interfaces separating the media, the contact lines near which strong manifestation of local system dynamics takes place: slipping, hysteresis of contact angles, etc. So, existence of hysteresis of a contact angle doesn't allow constructing the correct theory of parametric oscillations even in a linear approach [154] while non-linearity itself can be a reason for incorrectness of a statement of the boundary-value problem.

Davis and Weiland [167] showed that global instability of liquid films on inclined surfaces strongly depends on boundary conditions on the contact lines. Singularity of the tasks with existence of the interfacial and free boundaries evolving in space and time is caused by a microstructure of the contact lines and transition points, the location of which is changing in an unknown way.

In general, boundary conditions on dynamic interfaces of continua are non-linear and non-stationary. Non-linearity can be caused by kinematic (in case of free boundary) or by dynamic features: a contact angle hysteresis, the relaxation phenomena, a liquid separation from a wall in some places, etc. The conditions replacing classical conditions of sticking on a moving contact surface [168] are unknown; therefore it is impossible creating rather general adequate theory of parametric oscillations on the interfacial boundaries of continua.

### 2.5.3 Thermo-hydrodynamic processes on the interfacial boundaries

Features of thermo-hydrodynamic processes on the interfacial boundaries are defined by interaction of media in very thin layer having a thickness by order of radius of molecular interaction (the nano-sizes): $a' \sim 10^{-9}$ m. In this layer molecules' interaction in the phases causes that the physic-mechanical and chemical properties of the medium and its thermodynamic parameters differ from properties and parameters of the phases.

To study the specified phenomena the various approaches, for example, with introduction the $\Sigma$-phase concept [169] and consideration the micro-effects, as well as other methods [33,170-173] were implemented. The thermo-hydrodynamic equations for the boundary of continua may be written [169]:

$$\rho_j^0 (\vec{v}_j - \vec{v}_\Sigma) \cdot \vec{n} = idem, \qquad (15)$$

$$P_2 \cdot \vec{n} = P_1 \cdot \vec{n} + \rho_1^0 (\vec{v}_1 - \vec{v}_\Sigma) \cdot \vec{n} \cdot (\vec{v}_1 - \vec{v}_2) + \frac{1}{a'S_{12}} \oint_{\delta L'} \vec{p}_\Sigma(l')dl',$$

$$P_2 \cdot \vec{n} \cdot \vec{v}_2 = P_1 \cdot \vec{n} \cdot \vec{v}_1 + \frac{1}{a'S_{12}} \left[ \oint_{\delta L'} \vec{p}_\Sigma(l')\vec{v}_\Sigma dl' - \frac{d}{dt}a'e_\Sigma S_{12}\right] +$$

$$+\rho_1^0 (\vec{v}_1 - \vec{v}_\Sigma) \cdot \vec{n} \left[ e_1 - e_2 + \frac{|\vec{v}_1|^2 + |\vec{v}_2|^2}{2}\right] + (\vec{q}_1 - \vec{q}_2) \cdot \vec{n},$$

where $e_\Sigma, \vec{p}_\Sigma, \vec{v}_\Sigma$ are specific internal energy of $\Sigma$-phase (relating to the area of interfacial surface), surface tension force and velocity of the separating boundary movement, correspondingly. Here $\delta L'$ is a contour covering $S_{12}$ surface. The equations (15) can be used as boundary conditions for the equations of micro-movement in a system and for derivation of the macroscopic boundary conditions.

It should be noted that surface tension force depends on a surface tension coefficient defined for any two media by the temperature of two media and concentration of the containing impurity. Thus, on the boundaries separating solid and liquid or gaseous media the attraction between molecules of the solid medium and liquid (gas) considerably surpasses that between molecules of liquid (gas) owing to what in the majority of practically important cases rather good approximation to a real physical situation is the classical sticking condition.

### 2.5.4 Macroscopic boundary conditions

At a statement of macroscopic boundary conditions usually the phenomenological approach is used, as well as physical reasons and empirical laws. For example, considering the volume $V$ limited by the surfaces $S_1$ and $S_2$ parallel to a boundary interface $S_{12}$ of the media it is possible to choose the coordinate system in which $S_{12}$ is immovable. Then, directing distance between surfaces $S_1$ and $S_2$ to zero and neglecting influence of a transitional layer (a contact surface), from the differential equation array (5), (6) the conditions on the boundary interface are:

$$\rho_1 \vec{v}_1 \cdot \vec{n} \vec{v}_1 = (p_1 - p_2)\vec{n} + (\rho_1 - \rho_2)\vec{g} + (\rho_{e1}\vec{E}_1 - \rho_{e2}\vec{E}_2) \cdot \vec{n} +$$

$$+\rho_2 \vec{v}_2 \vec{n} \cdot \vec{v}_2 + (\vec{j}_1 \times \vec{B}_1 - \vec{j}_2 \times \vec{B}_2) \cdot \vec{n} + p_s^{ex}, \quad \rho_j \vec{v}_j \vec{n} = idem,$$

$$\vec{n} \times (\vec{E}_1 - \vec{E}_2) = 0, \quad \vec{n} \times (\vec{H}_1 - \vec{H}_2) = \vec{j}_s \times \vec{n},$$

$$\vec{n} \cdot (\vec{B}_1 - \vec{B}_2) = 0, \quad (\vec{D}_1 - \vec{D}_2) \cdot \vec{n} = \rho_{es}, \qquad (16)$$

$$\rho_1 \vec{v}_1 \cdot \vec{n} \left( e_1 + \frac{1}{2}|\vec{v}_1|^2\right) + (p_2 \vec{v}_2 - p_1 \vec{v}_1) \cdot \vec{n} + (\vec{q}_1^{ex} - \vec{q}_2^{ex}) \cdot \vec{n} = q_s^{ex} +$$

$$+\rho_2 \vec{v}_2 \cdot \vec{n} \left( e_2 + \frac{1}{2}|\vec{v}_2|^2\right) + (\vec{E}_2 \times \vec{H}_2 - \vec{E}_1 \times \vec{H}_1) \cdot \vec{n} + (\vec{q}_2^{in} - \vec{q}_1^{in}) \cdot \vec{n}.$$

Here $p_s^{ex}$, $q_s^{ex}$ - the surface densities of the external forces and energy fluxes' distributions, $\vec{j}_s, \rho_{es}$ are the surface current and charges' density.

Conditions (16) represent the mass, impulse, energy conservation on the boundary separating continuous media and are the main by consideration of parametric oscillations on the boundaries. The energy fluxes $\vec{q}$ and the fluxes of energy density $\vec{q}_s$ have non-thermal, non-mechanical and non-electromagnetic character (they are accounted in the other terms). The same is concerning to $p_s^{ex}$.

These conditions and (15) have quite general view; they are used as the corresponding starting boundary conditions by their statement in the specific physical situations. So, we mainly consider such tasks, for which superficial current and density of charges can be treated as negligible.

Besides, the other conditions reflecting the specifics of a concrete physical situation can be satisfied too: equality of temperatures, potentials (hydrodynamic, chemical, etc.). For the models of heterogeneous media sometimes it is necessary to use the stochastic boundary conditions too [159] if the processes are substantially stochastic ones.

## 3 Parametric oscillations on interfaces of the electro-conductive liquids

### 3.1 Features of the boundary interfaces

Behaviors of the boundary interfaces are defined by properties of media, conditions of their interaction and the nature of external influences (for example, electromagnetic - in case of the conductive media).

Because in complex heterogeneous media it is impossible to follow an evolution of each separate boundary of the phase separation, the various averaged equations and models obtained at the accounting of the media interaction on separate boundaries are used [25,169,172,174-176]. For example, stabilization systems for the boundaries of phase transition (crystallization) are based on use of high-frequent electromagnetic fields and active four-pole circuits with thermo-resistors and the corresponding operational amplifiers. Upon transition from a liquid to solid phase there is a jump in conductivity of media, which gives a chance for effective control of the crystallizing surface form applying the high-frequent electromagnetic fields.

### 3.1.1 Control systems and methods
Control systems for the phase transition boundaries built on the specified principle work as follows. Perturbation of the boundary of phase transition leads to the corresponding perturbation of a magnetic field, and a change of current in a winding of the operating control system caused by it, after corresponding strengthening, induce a secondary current in thin skin-layer near interfacial boundary of the phase separation, and by that promotes its alignment (stabilization) by the Joule heat fluxes.

Various features of the boundary control for the conductive liquids depending on specific physical situations were considered in a number of works:
- excitation of regular oscillations of interface by means of constant electric field of high intensity creating electric breakdown of a liquid [177];
- parametric impact on interfacial boundaries by crossed electric and magnetic fields [18,22,42];
- excitation of Kelvin-Helmholtz instability in alternating magnetic fields [123,178], Rayleigh-Taylor instability [87,98] and many other cases of excitation of the surface waves on interfacial boundaries [21,46,47,52,73,179].

### 3.1.2 Small magnetic Reynolds numbers
Mainly the regularities of excitation and suppression of the oscillations on the interfacial boundaries of conductive and non-conductive media by means of electromagnetic progressive or standing waves were of interest for us, by small magnetic Reynolds numbers $Re_m$, when the induced electromagnetic field is negligible in comparison to external field. Mainly the thin films of the conductive liquid possessing a number of surprising features in the theoretical and applied aspect were considered.

Due to the fact that the behavior of system is defined by parameters of external influences and Eigen oscillations, the task of parametric (electromagnetic in this case) excitation (suppression) of oscillations on the interfacial boundaries is formulated as a problem of determination and selecting the physically realized electromagnetic fields creating the requested effect: stabilization (suppression of instability) on boundaries of phase transition (crystallization) against casual or regular perturbations [21,22,32, 109,180-182], excitation of oscillations of the film surface in a requested form [4,6,9,46-49,87,134-137] and their disintegration into the drops (spraying, dispersing) [34,41,46,73,133,135,138-141,183]. Force and heat (Joule thermal emissions) influences of the electromagnetic fields on the media have been investigated.

### 3.1.3 Complex processes and singularities
The thermo-hydrodynamic processes on interfacial boundaries are distinguished as highly complex and multi-factored. For a number of physical situations they have been clarified [33,184-188]. In the majority of works the system of two phases separated by a layer of interfacial transition ($\Sigma$-phase) is replaced with a model zero thickness interface, on both sides of which the media are uniform by their properties up to the boundary.

Two parameters [167,189] appear in the general formulation of tasks with the separating boundaries: capillary number Ca=We/Re and the sliding coefficient characterized by the slipping length relation to linear scale of system. At $\sigma \to 0, Ca \to \infty$ ($\sigma$ is surface tension coefficient) there is a singularity in the boundary conditions (16). Thus, for Ca<<1 an interface is defined not by viscous deformation but by local dynamics: configurations of a local viscous flow, sliding of phases, etc.

Singularity in many cases yields a condition of sticking (non-slipping). These features lead to that in some physical situations the small changes of conditions at the interfacial boundary strongly influence to a change of a boundary form. So, for systems with a hysteresis of contact angle it is impossible to construct the correct stability theory even in a linear approach due to non-linearity of the boundary conditions [190] and strong sensitivity of global properties of parametric oscillations on the boundary interfaces to the conditions of the media contacts [169].

Due to specified features an interaction of electromagnetic fields with interfacial boundaries of conductive media in many aspects is not fully understood. But in connection with various requirements of practice a set of the special cases was considered giving sometimes the contradictive results not allowing creating integral understanding of the main regularities of phenomena. It testifies to great importance of the tasks, which solution is still found in an initial stage, despite abundance of separate publications with achievements in partly understanding of some phenomena [16,18,23,34,48, 49,73,109,117,133-137,155,178,191]. For example [183] it is claimed that the constant field suppresses instability of a tangential rupture more effectively than variable field, in [53] parametric effects aren't considered. The conclusion is drawn that the alternating electromagnetic field always destabilizes the interface boundary.

It was shown that alternating electromagnetic field stabilizes long-wave part of a spectrum of the two-dimensional perturbations of interface of the conductive and non-conductive liquids and causes progressing of instability having character of the parametric resonance [86]. The last one is very important, for instance, for the successful solution of the problem of a film MHD-granulation of metals [73,133,138-142] studied by us.

## 3.2 Boundary interfaces' parametric control
One of the first works on electromagnetic impact on a liquid metal jet with a purpose of its stabilization showed [21] that with a frequency of field satisfying the condition that a thickness of skin-layer is small comparing to capillary radius, the oscillations of a jet surface fade (steady jet). For electromagnetic levitation of liquid metals it was offered to use the law-frequency field of circular polarization for a jet's support in gravitation field and the high-frequency field for stabilization of instabilities.

### 3.2.1 Electromagnetic control of instabilities
Study of this problem was continued by many scientists. It was stimulated by need of stabilization the Rayleigh-Taylor [16,97,193], Kelvin-Helmholtz and Tonks-Frenkel instability [123,192,194,195]. The special kind instability of thin viscous jets and films, appearing the teethes on an interface (so-called "rugosity") was investigated on the one-dimensional and two-dimensional mathematical models [196], which analysis led to obtaining the conditions of appearing this type of instability connected to hyperbolic properties of the equations.

A number of the new theoretical and experimental works on parametric excitation of oscillations on interfaces of the conductive and non-conductive liquids by means of electromagnetic fields were considered in the reviews [35,197] containing also discussion on mechanisms of the fields-media interaction and analysis of quantitative and qualitative features of instability development and stabilization of the interfacial boundaries.

Stabilization of the free and surrounded liquid metal jets and films have been considered in touch with the thermonuclear technology [191,198,199], excitation of parametric instability, disintegration of the jet and film flows - in connection with a task of a granulation of metals [34,46,73,133].

### 3.2.2 Control of thermo-hydrodynamic processes
Problems of the thermo-hydrodynamic processes' control on the interfacial boundaries of phases in heterogeneous systems were considered for a case of the moving magnetized liquids in deformable [200] and non-deformable media [201,202]. In case of non-conductive liquid the control is possible by change of thermo-hydrodynamic parameters on the boundary (boundary control), for the conductive granular medium a current through the particles (heating in the points of a contact of granules owing to local increase in resistance leads to possibility of the volumetric control), for a flow of the conductive liquid the flow control is possible with non-conductive particles through using the special electromagnetic fields [203], etc.

In general the problem of parametric excitation and suppression of oscillations on the boundaries of the conductive liquid having important applications is at initial stage of development and the presented here results have shown an attempt of solution of a class of the tasks connected with disintegration and stabilization the liquid metal films and fronts of crystallization by means of progressive and standing electromagnetic waves.

Automatic control (regulation) of the processes in continua can be programmed [24] or adaptive (with feedback) [32,20-22,180] that is especially effective for the fast-proceeding and unstable processes. For this reason the tasks considered by us are important in the theoretical and applied aspect, especially for those technological processes and devices, which are under industrial demand. Classification of the studied physical processes of parametric excitation and suppression of oscillations on the interfaces of continua can be tracked visually according to the Table.

## 4 Oscillations in non-conductive media
Parametric excitation and suppression of oscillations on the interfaces of non-conductive media is possible due to action in the volumes occupied by media the distributed power sources or an exchange of mass, impulse and energy on the boundaries. It allows influencing effectively the kinematic and thermodynamic properties of media making the control of the processes running in media: stabilize interfaces of media in case of their instability against casual or regular perturbations [32,39,67,86,109,182], excite oscillations of the required mode [5,6,9,47,49,87], destroy the interfacial boundaries by excitation of the oscillations growing by amplitude (in time or by any coordinate) [26,34,35,41,46,73,116,204].

Table. Classification of considered problems

| Process | Electromagnetic control of oscillations boundaries of conductive media | Vibration control of jet and film flow decay | Thermal-dynamic oscillations on the boundaries of continua |
|---|---|---|---|
| Type of boundaries | Conductive liquid– non-conductive medium (liquid, gas, solid phase) | Liquid-liquid, gas | Solid phase - gas, liquid phase, boundaries of phase transition from liquid to solid state |
| Type of perturbation, medium | Progressive and standing electro-magnetic waves, Joule heat, homo-geneous hetero-geneous media | Vibration actions  Homogeneous media | Electromagnetic fields, heat and mass fluxes on interfacial boundaries, homogeneous and  heterogeneous media |
| Type of flow, phenomenon | Flat and radial spreading film flows, bending waves, film flow delay into drops | radial spreading film flows, bending waves, film flow delay into drops | Thermal-hydrodynamic oscillations granular media and on the crystallization boundaries |
| Main forces | Electromagnetic, capillary, gravitational, viscous | Vibration, capillary, gravitational, viscous | Thermodynamic, gravitational, viscous, interphase – heterogeneous medium |
| Non-linearity of processes and media | Non-linearity of waves, dispersion of surface and electromagnetic waves | Solitons and shock-waves, resonant pulverization of film flows | Non-linear thermo-hydrodynamic waves, non-linear physical characteristics of media |

## 4.1 Parametric oscillations of the interfaces

Anyway initiation of parametric fluctuations in continua requires exceeding some barrier of parameters determined by energy brought to system from the outside spent not only for a rating of fluctuations, but also for dissipation. As in the nature there are no absolutely elastic media, this lower barrier of parameters is defined by the dissipation energy.

Parametric excitation of oscillations on the boundary interfaces of continua [64,107,205] of infinite sequence of areas of unstable fluctuations gives the maximum width for the area of frequencies $\omega/2$ equal to a half of frequency of the compelling force (the main harmonica $\omega$).

From a set of various ways of excitation and suppression of the oscillations on boundaries of non-conductive media the acoustic ones [51,65,72], vibration and vibro-thermal [41,64], mechanical and gas-dynamic [45,206,207], are most often used, with application of the surface-active substances (SAS) [51,208], thermo-capillary effect [52,208-211] and others.

## 4.2 Transfer of wave energy between modes

In many works it was noted that parametric excitation (suppression) of some modes can lead to excitation (suppression) of the others. Parametric stabilization of some part of harmonics by the applied external influences can lead to excitation of instability by other kind or in other part of a range of frequencies [52,83].

In a majority of works the three main types of instability and possibility of their suppression by parametric action or their strengthening for the purpose of destruction of media are investigated: dispersing, spraying. One of the most developed areas - the theory of stability and disintegration of liquid jets and films - has numerous applications in various industrial, technical and other devices [34,73,90,135,212-214].

### 4.2.1 Special polyharmonic activators

Research of progressive the external perturbations of a jet's surface has shown that process is almost always generally defined by the first spectral mode. Complex wavy process is, as a rule, possible only by using special polyharmonic activators [77,215].

Resonant modes' interaction can lead to energy transfer from one mode to another. The phenomenon of generation the second harmonics called by Wilton's effect [216] is known, for example, when as a result of interaction of the gravitational and capillary waves with parameters $k_2 = k_1, \omega_2 = \omega_1$ at the flow velocities close to the threshold value, yields the strengthening of waves by half length (transfer of wave energy).

## 4.2.2 Parametric oscillations in film flows

The other surprising object of researches by parametric oscillations - film flows - differs in a hydrodynamic originality and broad practical application. It is a problem by stability and disintegration (dispersing, spraying) of the boundary interfaces and free surfaces, etc. [114-116,217-220]. The flows of thin liquid layers (films) are always unstable even in a linear approach if only lower of the adjoining liquids is more viscous [221], and non-linear saturation of instability in moving films takes place at the combined action of shift and a surface tension.

Attempts were made [121] to explain the general mechanism of the non-linear saturation of instability in thin films on example of the Rayleigh-Taylor instability. It was established that in the certain range of parameters' variation the perturbation breaks off a standing film and doesn't break the moving one.

The works [14,34,52,90,104,112,222] have been devoted to research by influence of external perturbations of unstable liquid films on dispersive structure of the drops which are formed as a result of their disintegration. So, the experimental study of the compelled high-frequency impact on stability of thin liquid films showed [223] its weak influence on increase rate of perturbations on the film surface, however influence on dispersive distribution of drops is considerable: reduction of the drops' sizes on average is 15-20%, and with some frequencies up to 40%.

In colloidal systems [224] spontaneous destruction of films was observed at a critical thickness. The phenomenon of a rupture is described on the basis of the numerical solution for a non-linear boundary task. Thus by a number of authors the need of the account also the thermo-capillary effects was indicated.

Influence of dynamic effect of air, disintegration of films in a direct-flow gas stream, fluctuations of a film surfaces, a shaping of the breaking-up boundary separating it from the surrounding medium, dispersing characteristics of a round film have been investigated in [86,90,104,222] and in other works. The behavior of interfacial boundaries of the films of incompressible liquid spreading in other liquid medium was studied in linear, as well as non-linear approaches.

Within the linear theory it was shown [225] that in a presence of negative tangent tension the bending perturbations of a film surface are progressing and provoke to surface oscillations without change a film thickness. Such fluctuations are called the bending ones in a contrast to the axisymmetric fluctuations connected with a change of a film thickness (symmetrical to middle surface).

It is known that long-wave film thickness perturbations accrue in case of prevalence of the negative stretching forces over capillary forces. Computing experiments on non-linear dynamics of free liquid films showed that parametric oscillations of interface with the liquid surrounding it amplify non-linear effects and lead to its disintegration.

The stability research of a two-dimensional viscous liquid flow conducted by Galerkin's method showed [222]: in a linear case stability is defined by Reynolds number Re whereas in a non-linear case there is still a stability threshold, which is correlated with the experimental data.

Vibration initiation of parametric oscillations on a surface of radially spreading water and solutions of polymers films was studied in [112]. Two types of bending perturbations of a film surface were theoretically considered: the running concentric waves created by the horizontal vibrating disk (vibrating in a vertical direction), which on a round jet is coming from the nozzle located over disk plane, and standing waves are formed from edges of a disk and from some defects on its surface if any.

Radial flows of thin liquid films, change of their thickness, velocity profiles, and pressure fluctuation on a wall, stability and transition of the laminar flow into a turbulent one were investigated by Japanese scientists [226]. Many other problems by parametric excitation and suppression of oscillations in films and jets of non-conductive liquid were considered in a number of reviews and monographs [21,33,44,90,170,181,214,227-229], analyzed vibration, acoustic, thermo-capillary, SAS, etc. external influences.

Thermo-hydrodynamic fluctuations in heterogeneous media and their influence on heat and mass exchange and hydrodynamic processes were investigated in [9,11,64,175,230]. So, the results of experimental researches of heat transfer coefficients and hydraulic resistance [230] (carried out at a stationary and pulsed gas flow in porous medium) showed increase of processes' intensity: with resonant frequencies the coefficients for the pulsing mode exceeded by 2-3 times the corresponding stationary values.

## 4.2.3 Industrial applications

Parametric excitation and suppression of oscillations on the interfacial boundaries of homogeneous and heterogeneous continua and their influence on intensity of technological processes is the new rapidly developing area of modern thermo-hydrodynamics and the control theory for processes with distributed parameters. It is the relatively new science promising great opportunities in creation of the new resource and energy saving highly effective technologies and economic high-productive devices [117,133,138-142], which are extremely needed for a modern industry and economy.

It should be noted that introduction of low-intensive perturbations usually rather leads to some decrease in indicators (not resonant case) while at high intensity of external influences the technological process is defined mainly by nature of these influences and practically doesn't depend on character of unperturbed state of system. The resonant effects allowing significantly increase an intensity of process or even to receive essentially new phenomenon at rather low power expenses are especially important for practical use [34,117,133, 135,142]. That opens direct ways to creation the new and to improvement the known technologies.

Physical and mathematical models of processes for cases of vibration and thermal influences were constructed, the computer programs were created by us, computational and physical experiments for detection of regularities by earlier unexplored physical systems were made [133,231-233]: excitation of wavy processes of the stated kind and disintegration of the film flows of viscous liquids into drops (dispersing, granulation), increase of intensity by thermo-hydrodynamic processes in the presence of parameters' oscillations on interfacial boundaries in heterogeneous continua and others.

Low-amplitude linear perturbations, as well as the non-linear ones (including high-amplitude) are considered, and problems are solved in a general statement covering the cases of excitation and suppression of parametric oscillations having specific features and characteristic scopes. Mathematical models have been constructed on modular type allowing solving the whole class of the tasks by parametric excitation and suppression of oscillations on the interfacial boundaries of non-conductive continua. They allowed adaptation on other relatives in the mathematical relation of a task with other external influences too.

The results obtained have shown the available practical conclusions and recommendations. New technological decisions, some concrete designs were provided and the information on their efficiency and practical tests has been supplied. Along with improvement of indicators by known technologies and devices also the essentially new technologies have been developed and the devices based on the beautiful, original hydrodynamic phenomena were constructed [138-141]. For example: dispersing of the liquid films on the vibrating disk at the Euler's numbers significantly exceeding unit [77,78,133], use of the resonant phenomena of electromagnetic film flow disintegration, the soliton-like modes of the film flow decay, etc.

The problem of parametric excitation and suppression of oscillations on the boundaries of continua is being solved by us in connection with the theory development and practical requirements. Generally for the development of new materials on the basis of granule technology, for which we created methods and devices by receiving particles of metals of a given size and with a high crystallization rate (cooling rate of drops was reached $10^4$ Celsius degrees per second!). These are so-called amorphous metals. The idea of their creation went from an assessment of durability of iron which Frenkel [188] gave at the beginning of the last century having specified that theoretical iron durability differs from the real one up to 1000 times.

## 5 Conclusions by the results obtained
The following conclusions have been made based on the results obtained:
- Parametric excitation of the jet and film flows' disintegration allowed inventing and successful constructing the new highly effective granulation and other machines implemented into practice of modern energy and new material science technologies
- Discovered and investigated by us three new phenomena on film flow disintegration may be of interest for spreading technologies and other applications
- Parametric excitation and suppression of the oscillations at the boundary interfaces in continua have been studied by many researchers and the results obtained revealed a number of useful practical application, e.g. stabilization of the processes, intensification, etc.
- Specific advantages of our methods, compared to other ones worldwide, consist in the results obtained on the subject considering practically complicated cases. We continue this activity taking into account the additional real physical properties of considered situations.
- Both analytical, as well as numerical methods were developed for solving the non-linear boundary problems, which present the new scientific direction in the field of control processes in continua.
- Developed experimental facilities allowed testing the revealed new phenomena and create

perspective technologies and devices for granulation of liquid metals.

*References:*
[1] Bishop R. *Vibration,* Cambridge: University Press, 2 ed., 1979, 176 p.
[2] Ganiev R.F., Lapchinski V.F. *Problems of mechanics in space technology,* M.: Mashinostroenie, 1978, 119 p. (In Russian).
[3] Glickman B.F. *Automatic control of liquid propellant rocket engines,* M.: Mashinostroenie, 1989, 272 p. (in Rus.).
[4] *The oscillatory phenomena in multiphase media and their use in technology*/ Ed. R.F. Ganiev, K.: Tekhnika, 1980, 142 p. (in Rus.).
[5] Novitskiy B.G. *Application of acoustic oscillations in chemical-technological processses,* M.: Chemistry, 1983, 192 p. (in Russian).
[6] Chandrasekhar S. *Hydrodynamic and hydromagnetic stability,* Oxford: Clarendon Press, 1961, 654 p.
[7] Glansdorf P. *Dynamical systems and microphysics,* Springer-Verlag, 1980, P. 199-224.
[8] Avduevskiy V.S., Barmin I.V., Grishin S.D., et al. *Problems of space industry*, M.: Mashinostroenie, 1980, 221 p. (In Russian).
[9] Basov N.I., Lyubartovich S.A., Lyubartovich V.A. *Vibro-forming of polymers*, Leningrad: Khimiya, 1979, 159 p. (In Russian).
[10] Brounshteyn B.I., Fishbein, G.A. *Hydrodynamics, heat and mass transfer in disperse systems*, L.: Khimiya, 1977, 280 p. (In Rus.).
[11] Galitseiskiy B.M., Ryzhov Yu.A., Yakush E.V., et al. *Heat and hydrodynamic processes in vibrating flows,* M.: Mashinostroenie, 1977, 256 p. (In Russian).
[12] Kirko I.M. *Liquid metal in electromagnetic field,* M.-L.: Energiya, 1964, 160 p. (in Rus.).
[13] Klopovskiy V.A. Research of acoustic disintegration of liquid on uniform drops and development of acoustic granulators for mineral fertilizers/ *Works VNII Chim. Mash*, M., 1975, № 71, P. 13-19 (in Russian).
[14] Nakoryakov V.E., Pokusaev B.G., Shreiber I.R. *Wave dynamics of gas- and vapor-liquid media,* New York: Begell House, 1992, 246 p.
[15] Berezin Yu.A. *Numerical study of non-linear waves in the rarefied plasma*, Novosibirsk: Nauka, 1977, 112 p. (In Russian).
[16] Buchin V.A., Tsypkin A.G. About Rayleigh-Taylor instability of polarized and magnetized liquids in electromagnetic field// *Reports of Acad. Sci. USSR. Doklady akademii nauk*, 1974, Vol. 219, № 5, P. 1085-1088 (In Rus.).
[17] Vladimirov V.V., Mosiyuk A.I. Parametric excitation of short capillary waves on the surface of a liquid metal contiguous with an unstable plasma// *Soviet Phys. JEPT.- Pis'ma Zh. Tekh. Fiz*, 1983, Vol. 9, № 4, P. 236–238.
[18] Gelfgat Yu.M., Lielausis O.A., Shcherbinin E.V. *Liquid Metal Subject to the Influence of Electromagnetic Forces*, Riga: Zinatne, 1976, 248 p. (in Russian).
[19] Kukhtenko A.I., Ladikov Yu.P. Theory of automatic regulation and thermonuclear syntesis/ *Reports of USSR-USA conf. On dynamic stabilization of plasma*, Sukhumi, 1975, P. 79.
[20] Ladikov Yu.P., Samoilenko Yu.I. Application of system of orthogonalized wires with automatically regulating currents for stabilization of plasma in systems tokamak// *J. Eng. Physics (Zhurnal tekhn. fiziki)*, 1972, Vol. 17, № 9, P. 129-134 (in Russian).
[21] Ladikov Yu.P. *Stabilization of processes in continua,* M.: Nauka, 1978, 432 p. (in Russian).
[22] Ladikov Yu.P., Tkachenko V.F. *Hydrodynamic instabilities in metallurgical processes*, M.: Nauka, 1983, 248 p. (in Russian).
[23] Cēbers A.O. Instability of the free surface of a magnetic fluid in a tangential rotating magnetic field// *J. Magnetohydrodynamics*, 1981, Vol. 17, № 3, P. 41-49; transl.: 1981, Vol. 17, № 3.
[24] Butkovsky A.G. *Optimal Control of Distributed Parameter Systems*, N.Y.: Elsevier, 1969, 474 p.
[25] Goldstick M.A. *Transfer processes in granular layer*, N-sk: In-t of thermal physics, Siberian branch Acad. Sci, 1984, 164 p. (In Russian).
[26] Kodzhaev Sh.Ya., Kochetkov A.A. Experimental study of mechanism of cyclic oil recovery from jointed and porous media, M.: Nauka, 1970, 140 p. (in Russian).
[27] Zabarnyi G.N. *Calculation methods for processses of heat transfer in systems of extraction of geothermal energy*/ Dis. D.Sc., Ukr. Acad. Sci., In-t energy saving problems, Kiev, 1995, 356 p.
[28] Sardarov S.S. (Jr.), Savina E.V. Geothermal field of a break// *Reports of Acad. Sci. USSR.- Doklady akademii nauk,* 1984, Vol. 276, № 5, P. 1091-1094 (In Russian).
[29] Davydov A.S. *Solitons in Molecular Systems*, Reidel, Dordrecht, 1985, 288 p.
[30] *Mathematical modeling. Processes in non-linear media,* M.: Nauka, 1986, 312p. (in Rus.).
[31] Lomdahl P.S., Layne S.P., Bigio I.J. Solitons in biology// *Los Alamos Sci,* 1984, № 10, P. 2-22.
[32] Vasiliev V.I., Ladikov Yu.P., Kazachkov I.V. *About stability and stabilization of garnissage*

*in steel-smelting machines*, K., 1981, 33 p. (preprint/ In-t Cybernetics Acad. Sci., №81-18.) (In Russian).

[33] *Hydrodynamics of interphase surfaces*. In: Collection of Papers, M.: Mir, 1984, 210 p.

[34] Kolesnichenko A.F. *Technological MHD-devices and processes*, Kiev: Nauk. Dumka, 1980, 192 p. (in Russian).

[35] Nevolin V.G. Parametric excitation of surface waves// *J. Engineering Physics and Thermophysics*, Transl. from IFZh, 1984, Vol. 47, № 6, P. 1028–1042.

[36] Chernov D.K. *Notes of Russian engineering society*, S.-Petersburg, 1879, Issue I, P. 140-152 (In Russian).

[37] Potapevskiy A.G. Lapchinskiy V.F., Vainerman A.E. *Pulse-arc welding of aluminum alloys*, L.: Znanie, 1966, P. 38-43 (in Russian).

[38] Slavin G.A., Stolpner E.A., et al. Control of crystallization process through dynamic arc action// *Welding production*, 1974, № 8, P. 23 (In Russian).

[39] Kazachkov I.V. Stability of slag lining in steel-smelting units// *Soviet automatic control*, 1982, Vol. 15, P. 74-81.

[40] Lions J.-L. *Optimal control of systems governed by partial differential equations*, Berlin: Springer Verlag, 1971, 396 p.

[41] Ageev S.G. Condition of violation of a continuity of a free surface of the vibrating liquid column. In: *Optimization of parameters of machines and industrial processes. Scientific works of ChPI*, № 180, Chelyabinsk, 1976, P. 24-28 (In Russian).

[42] Briskman V.A. and Shaidurov G.F. Parametric excitation of fluid instability in magnetic and electric fields// *Magnetohydrodynamics*, 1969, Vol. 5, № 3, P. 15-19; transl.: V. 5, №3, 10-12.

[43] Grigoryan S.S., Zhygachev L.I., Kogarko B.S., et al. Parametric resonance in communicating vessels at vertical variable loadings// *J. Fluid dynamics. Transl. Izvestiya Akad. Nauk, mekhanika zhidkosti i gaza*, 1969, № 2, P. 148-150.

[44] Dityakin Yu.F., Klyachko L.A., Novikov B.F., et al. *Dispersion of liquids*, M.: Mashinostroenie, 1977, 208 p. (In Russian).

[45] Entov V.M., Sultanov F.M., Yarin A.L. Disintegration of liquid films subjected to an ambient gas pressure difference// *Fluid dynamics*, Springer, 1986, 21, S. 376-383.

[46] Kazachkov I.V., Kolesnichenko A.F. Decay of liquid metal films under effect of electromagnetic field// *Engineering electrodynamics*, 1984, № 3, P. 16-19 (in Russian).

[47] Kazachkov I.V. Electromagnetic wave excitation in a liquid film// *Magnetohydrodynamics*, 1985, v. 21, № 3, P. 273-279. Transl.: Magnit. Gidrodinamika, 1985, № 3, P.77-82.

[48] Kazachkov I.V., Kolesnichenko A.F. About disintegration of liquid-metal film in non-conducting medium with electromagnetic field. *Magnetohydrodynamics*, 1984, № 1, P. 44-46.

[49] Kazachkov I.V. The influence of an alternating electromagnetic field on the wave spreading out in a liquid metal film// *Magnitnaya Gidrodinamika (Magnetohydrodynamics)*, 1985, № 3, P. 135-137 (in Russian).

[50] Nevolin V.G. Parametric instability of liquid films. In: *II USSR seminar on hydromech. and heat mass transfer in zero gravity/* Abstr.-Perm. In-t mech. of continuous media, Ural Acad. Sci, 1981, P. 117,118 (in Russian).

[51] Nevolin V.G. Influence of soluble surfactants on liquid dispergating// *Izv. Akad. Nauk SSSR, Mekhan. Zhidk. Gaza*, 1981, № 5, P. 160-164.

[52] Nevolin V.G. Influence of thermocapillary effect on parametric excitation of surface waves// *J. Engineering Physics and Thermophysics (IFZh)*, 1984, № 3, P. 424-428 (Rus).

[53] Garnier M. Ro´le de´stabilisant d´un champ magnetique alternatif applique´ au voisinage d´une interface// *C.R. Acad. Sci. Paris*, 1977, Se´r B, v. 284, P. 365-368.

[54] Verte L.A. *Electromagnetic transport of liquid metals*, M.: Metallurgiya, 1965, 236p. (In Rus).

[55] Povkh I.L., Kapusta A.B., Chekin B.V. *Magnetohydrodynamics in metallurgy*, M.: Metallurgy, 1974, 240 p. (in Russian).

[56] Karpacheva S.M., Raginskiy L.S., Muratov V.M. *Bases of the theory and calculation of horizontal pulsation devices and pulsators*, M.: Atomizdat, 1981, 191 p. (in Russian).

[57] Kafarov V.V. *Cybernetic methods in chemistry and chemical engineering*, Central Books Ltd, 1977, 484 p.

[58] Rauschenbach B.V. *Vibrational combustion*, M.: Fizmatgiz, 1960, 500 p. (in Russian).

[59] Kurdyumov S.P. *Localization of diffusive processes and appearance of structures by development in the dissipative medium the modes with an aggravation/* Dis. D.Sc., M.: In-t Applied math. Acad. Sci. USSR, 1979 (in Rus).

[60] Martinson L.K. Spatial localization of heat perturbations in moving non-linear media// *High-temperature thermal physics (TVT)*, 1979, № 5, P. 1019-1023 (in Russian).

[61] Samarskii A.A., Galaktionov V.A., Kurdyumov S.P., Mikhailov A.P. *Action of the limit modes with an aggravation on medium with constant*

*heat conductivity*, M.: Preprint/IPM AN SSSR, №28, 1979, 76 p. (in Russian).

[62] Samarskii A.A., Zmitrenko N.V., Kurdyumov S.P., et al. *Metastable heat localization in a medium with non-linear heat conductivity and conditions for its reveal in experiment*, M.: Preprint/IPM AN SSSR, № 103, 1977, 87 p. (in Russian).

[63] Samarskii A.A., Zmitrenko N.V., Kurdyumov S.P., et al. Effect of metastable heat localization in a medium with non-linear heat conductivity// *Reports of Acad. Sci. USSR, Doklady akademii nauk*, 1975, Vol. 223, № 6, P. 1344-1347.

[64] Astrakhan I.M., Gadiyev S.M., Trusfus A.V. Estimation of effectiveness of vibro-thermal impact on oil layer. *Works of Moscow In-t of oil chemistry and gas industry*, 1984, № 186, P. 45-50 (In Russian).

[65] Boguslavskiy V.Ya., Eknadiosyants O.K. On the physical mechanism of liquid spraying by acoustic oscillations// *Acoustical journal*, 1969, Vol. 15, № I, P. 17-25 (In Russian).

[66] Galaktionov V.A. Some properties of travelling waves in a medium with nonlinear thermal conductivity and a source of heat// *USSR Computational Mathematics and Mathematical Physics*, 1981, Vol. 21, № 3, P. 167–176.

[67] Nevolin V.G. Possible mechanism of acoustic foam clearing// *J. Eng. Physics (ZhTF)*, 1980, Vol. 50, № 7, P. 1556-1558 (in Russian).

[68] Minoshima W., White J.L. Stability of continuous film extrusion process// *Polym. Eng. Reviews*, 1983, Vol. 2, № 3, P. 212-226.

[69] Ponstain J. Instability of rotating cylindrical jets// *Appl. Scientific research*, 1959, A8, № 6, P. 425-456.

[70] Repin V.B. Heat transfer by flow of viscous fluid in flat channel under field of transversal acoustic oscillations// *Izv. Akad. Nauk SSSR, Mekhan. Zhidk. Gaza*, 1985, № 3, P. 41-49.

[71] *Physics and technique of powerful ultrasound. V.3. Physical basics of ultrasound techno-logy*/ Ed. L.D. Rosenberg, M.: Nauka, 1970, 380 p.

[72] Eisenmenger W. Dynamic properties of the surface tension of water and aqueous solutions of surface active agents with standing capillary waves// *Acoustics*, 1959, Vol. 9, P. 237-340.

[73] Kolesnichenko A.F., Kazachkov I.V., Vodyanyuk V.O., Lysak N.V. *Capillary MHD flows with free surfaces*, Kiev: Nauk. Dumka, 1988, 176 p. (In Russian ).

[74] Kozlov L.F. *Theoretical studies of boundary layers*, K.: Nauk. Dumka, 1982, 293p. (in Rus).

[75] Bussman K., Muenz H. Die stabilität der laminar Reibungsschicht mit Absaugung/ *Deutsch. Lutabfahrtforsch. Zbl. Wiss.*, Ber.-Wes, 1942, B. 1, S. 35.

[76] Wuest W. Nahrungsweise Berechnung und Stabilitätsverhalten von laminar Grenzschichten mit Absaugung durch Einzelschlitze// *Ingen. Arch*, 1953, Vol. 21, P. 90.

[77] Voloshyn A.Ya., Ivanochok I.M., Kazachkov I.V. *Electrodynamic excitation of the liquid film decay and development of the perspective MHD granulators*, Kiev: IED Acad. Sci., Preprint № 587, 38 p. (In Russian).

[78] Glukhenkiy A.I., Kazachkov I.V., Kolesnichenko A.F., et al. *Electromechanical control of the jets and films of viscous fluid and development of perspective granulators*, K.: 1988, preprint/ In-t Electrodynamics, №527, 40p. (In Russian).

[79] Ashby W.R. *An Introduction to Cybernetics*, London: Chapman & Hall, 1957, 294 p.

[80] Butkovsky A.G., Poltavsky L.N. Optimal control of wave processes// *J. Automatics and telemechanics. Avtomatika i telemekhanika*, 1966, Vol. 27, № 9, P. 30-34 (In Russian).

[81] Sirazetdinov T.K. *Stability of systems with distributed parameters*, N.: Nauka, 1987, 231p.

[82] Sirazetdinov T.K. To the analytical construction of the regulators for magnetohydrodynamic processes// *J. Automatics and telemechanics (zhurnal Avtomatika i telemekhanika)*, 1967, Part 1, П, №№ 10, 12 (In Russian).

[83] Samoilenko Yu.I. About resolution ability of the spatially distributed control systems// *Izv. AN SSSR. Techn. Cybernetics*, 1966, № 4, P. 137-142 (In Russian).

[84] Butkovskiy A.G., Samoilenko Yu.I. *Control of quantum-mechanical processes and systems*, Dordrecht: Kluwer Acad. Publ, 1990, 232 p.

[85] *Hydromechanics and transfer processes in zero gravity*, Sverdlovsk: Ural scientific center of Acad. Sci. USSR, 1983, 168 p. (In Russian).

[86] Korovin V.M. Rayleigh-Taylor instability of surface separating conductive and non-conductive fluid in alternating magnetic field// *Izv. Akad. Nauk SSSR, Mekhan. Zhidk. Gaza*, 1983, № 1, P. 31-37 (in Russian).

[87] Peskin R., Raco R. Ultrasonic atomization of liquids// *J. Acoust. Soc. Am*, 1963, Vol. 35, № 9, P. 1378-1385.

[88] Afraymovich V.S., Bykov V.V., Shilnikov L.P. About emergence and structure of the Lorentz attractor, *Reports of Acad. Sci., Doklady akademii nauk SSSR*, 1977, Vol. 234, № 2, P. 336-339 (In Russian).

[89] *Hydrodynamic and convective stability of incompressible fluid*, S.: Ural scientific center of Acad. Sci. USSR, 1984, 120p. (In Russian).

[90] Entov V.M. and Yarin A.L. Dynamics of Free Liquid Jets and Films of Viscous and Rheologically Complex Liquids, *Advances in Mechanics, VINITI, Mekhanika Zhidkosti i Gaza*, 1984, 18, P. 112-197 (in Russian).

[91] Arneodo A., Coullet P., Tresser C. A possible new mechanism for the onset of turbulence// *Phys. Letters*, 1981, Vol. 81A, №4, P. 197-201.

[92] Collet P., Eckmann J.-P., Landford O.E. Universal properties of maps on the interval// *Commun. Math. Phys*, 1980, Vol. 76, № 3, P. 211-254.

[93] Feigenbaum M.J. Quantitative universality for a class of nonlinear transformation// *J. Stat. Phys*, 1978, Vol. 19, № 1, P. 25-51.

[94] Aslanov S.K., Borko V.P. Study of temperature perturbations and their action on stability of flow in round tube with varying viscosity. In: *Mathematical methods of heat and mass transfer*, D-sk: DGU, 1980, P. 71-77 (In Russian).

[95] Belotserkovskiy O.M. *Numerical simulation in mechanics of continua*, M.: Nauka, 1984, 519p.

[96] Bird R.B., Stewart W.E., Lightfoot E.N. *Transport phenomena*, J. Wiley&Sons, 2007, 905p.

[97] Koroteev A.S., Rei I.N. Rayleigh-Taylor instability in thin liquid metal films by presence of magnetic field// *Izv. Sibir. Akad. Nauk SSSR, ser.tekh. n.*, 1984, № 10, № 2, P. 113-116.

[98] Lin S.P. Effects of Surface Solidification on the Stability of Multi-Layered Liquid Films// *ASME J. Fluids Eng*, 1983, 105, №1, P.119-120.

[99] Martinson L.K. Evolution of thermal impulse in medium with non-linear heat conductivity, *Works of Moscow Bauman university (MVTU)*, 1982, № 374, P. 14-34 (in Russian).

[100] Samarskii A.A., Sobol I.M. Examples of numerical calculation of temperature waves// *J. Comp. Math. and Math. Physics, ZhVMMF*, 1963, Vol. 3, № 4, P. 702-719 (In Russian).

[101] Samarskii A.A. *The Theory of Difference Schemes*, USA: Marcel Dekker, 2001, 788 p.

[102] Batchelor G.K., Moffatt H.K., Worster M.G. (eds), *Perspectives in Fluid Dynamics*, Cambridge: U.P., 2008, 501 p.

[103] Ciliberto S., Gollub J.P. Chaotic mode competition in parametrically forced surface waves// *J. Fluid Mech*, 1985, 158, P. 381-398.

[104] Nobuki, Mori Kazutoshi, Kimura Nasahiro et al. Study on disintegration of liquid film by air flow// *Trans. Jap. Soc. Mech. Eng.*, 1985, B51, № 469, P. 2880-2886.

[105] Arbuzov V.A., Kuznetsov E.A., Noskov N.N., et al. *On parametric excitation of waves on surface of fluid*, N.: 1977, 13 p. (In-t automation and electromeasurements, Sib. B. Acad. Sci., preprint №57) (In Russian).

[106] Bernshtam V.A., Kozyrev S.V., Kostenko Yu.T., et al. Liquid-metal protection of the elements of perspective energy installations, Kh.: *NTO energetiki i elektrotekhn. promyshl.*, 1987, 80 p. (In Russian).

[107] Velikhov E.P. The stability of a plane Poiseuille flow of an ideally conducting fluid in a longitudinal magnetic field// *Soviet Phys. JEPT*, 1959, Vol. 9, P. 848.

[108] Gavrikov V.K., Kats A.V., Kantorovich V.M. The forced dispersion on superficial waves// *Reports of Acad. Sci. USSR, Dokl. Akad. nauk*, 1969, Vol. 186, №5, P. 1052-1054 (In Rus.).

[109] Gagarin A.G. Influence of constant electric field on a film flow of liquid dielectric// *J. of Eng. Physics and Thermo-physics. IFZh*, 1985, Vol. 48, № 3, P. 18-22.

[110] Gogosov V.V., Naletova V.A., Taktarov N.G., et al. Hydrodynamics of surface phenolmena// *J. Applied Math. And Mech., Prikl. Matem. i mekh.*, 1984, 48, № 3.- P. 388-396.

[111] Entov V.M. *Dynamics of viscous and elastic fluid films*. Preprint № 130, Inst. Prikl. Mekhan. Akad. Nauk SSSR, M.: 1979, 47 p. (In Rus.).

[112] Entov V.M., Rozhkov A.N., Feizkhanov U.F., et al. Dynamics of liquid films. Plane films with free rims// *J. appl. Mech. and technical physics*, Springer, 1986, Vol. 27, S. 41-47.

[113] Entov V.M., Kestenboim Kh.S., Rozhkov A.N., et al. On the dynamic equilibrium mode of viscous and viscoelastic fluid films// *Izv. Akad. Nauk SSSR, Mekhan. Zhidk. Gaza*, 1980, №2, P. 9-18 (in Russian).

[114] Epikhin V.E. About the forms of ring jets of liquid// *Izv. Akad. Nauk SSSR, Mekhan. Zhidk. Gaza*, 1977, № 1, P. 9-14 (in Russian).

[115] Epikhin V.E. About the forms of ring jets of liquid// *Izv. Akad. Nauk SSSR, Mekh. Zhidk. Gaza*, 1979, № 5, P.144-148 (in Russian).

[116] Epikhin V.E., Shkodov V.Ya. About the forms of ring jets of liquid// *Izv. Akad. Nauk SSSR, Mekh. Zhidk. Gaza*, 1983, № 6, P. 3-11.

[117] Kolesnichenko A.F., Kazachkov I.V., Vodyanuk V.O., et al. *Development and research of construction of generator for jet-drop flow to diaphragm of experimental thermonuclear reactor*., K.-L.-M., IED-NIIEFA-Kurchatov in-t. Reports, 1986, 80 p.; 1987, 53 p.

[118] Kolovandin B.A. Stability of conductive fluid flow with free surface under magnetic and electric fields// *Appl. Mech. (Prikl. mekhanika)*, 1965, Vol. I, P. 95-105 (in Russian).

[119] Kolovandin B.A. Stationary flow of viscous conductive fluid with free surface under crossed electric and magnetic fields. In: *Engineering electromagnetic hydrodynamics*, Donetsk, 1965, P. 321-324 (in Russian).

[120] Kolodyazhniy A.A. *Solution's bifurcations for magneto-hydrodynamic equations in the problems with free surfaces*/ Diss. Cand. Phys.-math. Sci., K.: In-t Cybernetics, 1985, 117 p.

[121] Babchin A.J., Frenkel A.L., Levich B.G. Nonlinear saturation of Rayleigh-Taylor instability in thin films// *Phys. Fluids*, 1983, 26, № 11, P. 3159-3161.

[122] Gilmor R. *Catastrophe Theory for Scientists and Engineers*, Dover Publ., 1993, 666 p.

[123] Korovin V.M. Kelvin-Helmholtz instability in alternating magnetic field// *J. Appl. Mech. and Eng. Physics (ZHPMTF)*, 1984, № 2, P. 94-98.

[124] *Some problems of stability of a liquid surface*/ Ed. Briskman V.A. S-sk, Preprint In-t mech. cont. media, 1984, 77 p. (in Russian).

[125] Sprott J.C. *Strange Attractors: Creating Patterns in Chaos*, M&T Books, New York, 591 p.

[126] *Non-linear waves: self-organization*, M.: Nauka, 1983, 263 p. (in Russian).

[127] *Nonlinear Waves*/ S. Leibovich & A.R. Seebass, eds., Ithaca/London: Cornell U.P., 1974, XII + 331 p.

[128] *Nonlinear Waves: Structures and Bifurcations*/ Ed. A.V. Gaponov-Grekhov, M.I. Rabinovich, M.: Nauka, 1987, 401 p. (in Russian).

[129] *Non-linear wave processes in two-phase media*/ Materials of XX Siberian thermal phys. Seminar, Ed. S.S. Kutateladze, 1976, ITF SO AN SSSR, 1977, 417 p. (in Russian).

[130] Non-linear waves/ Materials of III sci. School on non-linear oscillations and waves in distributed systems, *Izv. vuzov. Radiofizika*, 1976, Vol. XIX, № 5,6 (in Russian).

[131] Korteweg D.J. Vries G. de. On the change of form of long waves advancing in a rectangular channel and on a new type of long stationary waves*// Philos. Mag.*, 1895, V.39, P.422- 443.

[132] Lavrentiev M.A. To the theory of long waves, *Works of In-t mathematics*, Ukr. Acad Sci., 1946, № 8, P. 13-69 (in Russian).

[133] Kazachkov I.V. *Parametric wave excitation and suppression on the interfaces of continua*/ D.Sc. thesis in Mechanical Engineering, Kiev, Institute of Electrodynamics, 1989, 464 p.

[134] Kazachkov I.V. Analytical solutions for electromagnetic excitation of nonlinear waves in films*, Magnetohydrodynamics,* vol.27, № 2; April-June 1991; p.190-5. Transl.: Magnitnaya-Gidrodinamika. vol.27, № 2, 1991, p. 83-89.

[135] Kazachkov I.V., Kolesnichenko A.F. About the phenomena of electromagnetic excitation of film decay, *Pis'ma v Zh. Tekhn. Fiziki (Letters to J. Techn. Phys.)*, 1991, V.17, № 3, P. 40-43.

[136] Kazachkov I.V. Electromagnetic wave excitation and suppression of films, *Magnetohydrodynamics,* vol.32, № 1, 1996; p.68-73. Trans.: Magnit. Gidrodin. 1996, № 1, p.74-80.

[137] Kazachkov I.V., Lysak N.V. FORTRAN codes for solving non-stationary boundary problems for parametric oscillations spreading in flows with free surfaces and saturated granular media. *Numerical Simulation of the Problems of Mathematical Physics*. Kr-sk St. Univ. 1988.

[138] Kazachkov I.V., Kolesnichenko A.F., Kuchinsky V.P., et al. *The device for powder produce from molten metal*. SU Patent, 1991.

[139] Vodyan'yuk V.O., Kazachkov I.V., Kolesnichenko A.F., et al. *The device for produce of spherical particles from molten metal*. SU Patent, 1990 (in Russian).

[140] Kazachkov I.V., Kolesnichenko A.F., Kuchinsky V.P., et al. *The method of powder' produce from molten metal*. SU Patent, 1989.

[141] Kazachkov I.V., Kolesnichenko A.F., Kuchinsky V.P., et al. *The method of particles' produce from molten metals and device for its realization*. SU Patent. 1988 (in Russian).

[142] Gorislavets Yu.M., Kazachkov I.V., Kolesnichenko A.F., et al. *Development of MHD method of Al-alloy granulation*. Report Inst. Electrodyn., Kiev, 1987, 132 p. (in Russian).

[143] Zel'dovich Ya.B. and Kompaneets A.S. The theory of the propagation of heat with a thermal conductivity depending on the temperature. In: *Collection Dedicated to the Seventieth Birthday of A.F. Ioffe*, M.: AN SSSR, 1950, P. 61-71.

[144] Samarskii A.A., Galaktionov V.A., Kurdyumov S.P., Mikhailov A.P. *Blow-Up in Quasilinear Parabolic Equations*, Berlin: Walter de Gruyter & Co, 1995, xxii+535 p.

[145] Martinson L.K. Thermal self-isolation of the localized structures in media with volumetric absorption of heat// *Letters to the J. Eng. physics (Pis'ma v zhurnal tekhn. fiziki)*, 1980, Vol. 6, № 4, P. 211-215 (in Russian).

[146] Balescu Radu. *Statistical mechanics of charged particles*, N.Y.: Intersci. Publ, 1963, 465 p.

[147] Bogoliubov N.N. *Problems of Dynamic Theory in Statistical Physics*, Oak Ridge, Tenn.: Technical Information Service, 1960, 120 p.

[148] Gyarmati I. *Non-equilibrium Thermodynamics. Field Theory and Variation Principles*, transl. by E. Gyarmati and W.F. Heinz, Berlin: Springer, 1970, 304 p.

[149] Zak M.A. *Non-classical problems in continuum mechanics*, L.: LGU, 1974, 120 p.
[150] Yavorsky N.I. Variation principle for viscous heat conductive liquid with relaxation// *Izv. Acad. Sci. USSR. Mech. zhidk. i gaza*, 1986, № 3, P. 3-10 (In Russian).
[151] Benoit B. Mandelbrot. *The Fractal Geometry of Nature*, S. Francisco: Freeman, 1983, 480 p.
[152] Samko S., Kilbas A.A. and Marichev O. *Fractional Integrals and Derivatives: Theory and Applications*, Taylor&Francis, 1993, 1006p.
[153] Babenko Yu.I. *Heat-Mass Transfer: Methods for Calculation of Thermal and Diffusional Fluxes*, L.: Khimiya, 1986, 236 p. (In Rus.).
[154] Sedov L.I. *Mechanics of continuous media*, USA: World Scientific; 4$^{th}$ ed. 1996, 1368 p.
[155] Davis S.H. Moving Contact Lines and Rivulet Instabilities. Part 1. The Static Rivulet// *J. Fluid Mech*, 1980, 98, P. 225-242.
[156] Roache P.J. *Computational fluid dynamics*, Albuquerque: Hermosa, 1976, VII+ 446 p.
[157] Tikhonov A.N., Samarskii A.A. Equations of mathematical physics, N.Y.:Dover, 1990, 724p.
[158] Richardson L.F. The approximate arithmetical solution by finite differences of physical problems involving differential equations, with application to the stresses in a masonry dam// *Trans. Roy. Soc. London*, 1910, Ser.A, V. 210, P. 307-357.
[159] Goritskiy S.V. About statement of stochastic boundary conditions in numerical model of granular medium, *Problems applied mechanics*, M.: MFTI, 1986, P. 49-54 (In Russian).
[160] Dorodnitsyn A.A. Computational methods in physics and mechanics: problems and perspective/ *Proc. 4$^{th}$ Austral. comput. conf.* Adelaide: South Australia, 1969, P. 667-671.
[161] Cheng S.I. Numerical integration of Navier-Stokes equations// *AIAA Journal*, 1970, Vol. 8, № 12, P. 2115-2122.
[162] Allen D.N. de G., Southwell R.V. Relaxation methods applied to determine the motion in two dimensions of a viscous fluid past a fived cylinder// *Quart. J. of Mech. and Appl. Math*, 1955, Vol. 8, P. 129-145.
[163] Moretti G. The importance of boundary conditions in the numerical treatment of hyperbolic equations/ *Polytechn. Inst. Brooklyn, PIBAL Rept*, 1969, № 68, P. 34.
[164] Thom A., Apelt C.J. *Field computations in engineering and physics*, London, N.Y.: C. Van Nostrand Company, Ltd, 1961, 168 pp.
[165] Thoman D.C., Szewczyk A.A. *Numerical solutions of time dependent two dimensional flow of a viscous, incompressible fluid over stationary and rotating cylinders*, Techn. Rept. 66-14, Notre Dame, Indiana.
[166] Alexandrov V.M., Kovalenko E.V. *Problems with mixed boundary conditions in continuum mechanics*, M.: Nauka, 1986, 336 p. (In Rus.).
[167] Weiland R.H., Davis S.H. Moving Contact Lines and Rivulet Instabilities. Part 2. Long Waves on Flat Rivulets// *J. Fluid Mech*, 1981, Vol. 107, P. 261-280.
[168] Dussan V E.B. Hydrodynamic stability and instability of fluid systems with interfaces// *Arch. Rat. Mech. Anal*, 1975, Vol. 57, P. 363.
[169] Nigmatulin R.I. *Fundamentals of Mechanics of Heterogeneous Media*, M.: Nauka, 1978, 336p.
[170] Kutateladze S.S., Nakoryakov V.E. *Heat and Mass Transfer and Waves in Gas-Liquid Systems*, N-sk: Nauka, 1984, 304 p. (in Russian).
[171] Nigmatulin R.I. Dynamics of Multiphase Media, USA: Hemisphere, 1991, 1, 507p.; 2, 371p.
[172] Nikolaevskiy V.N., Basniev K.S., Gorbunov A.T., Zotov G.A. Mechanics of Saturated Porous Media, M.: Nedra, 1970, 335 p. (in Russian).
[173] Nikolaevskiy V.N. Mechanics of jointed and porous media, M.: Nedra, 1984, 232 p.
[174] Bakhvalov N.S., Panasenko G.N. *Averaging of processes in periodic media*, M.: Nauka, 1984, 352 p. (In Russian).
[175] Sanchez-Palencia E. *Non-Homogeneous Media and Vibration Theory*, Springer-Verlag, 1980, 398 p.
[176] *Free Boundary Problems: Applications and Theory*, Vol. 3/ Ed. Bossavit A, Boston: Pitman, 1985, 303 p.
[177] Nesterov S.V., Sekerzh-Zenkovich S.Ya. Self-oscillations of the non-uniform liquid placed in electric field// *Reports of Acad. Sci., Dokl. Akad. nauk*, 1981, 256, №2, P.318-320
[178] Lindsay K.A., Kelvin-Helmholtz instability for a viscous interface// *Acta Mech.*, 1984, Vol. 52, № 1-2, P. 51-61.
[179] Saha Susama, Chaudhuri Krishna M. Thermal convection instability of liquid metals in magneto-hydrodynamics// *Astrophys. and Space Sci*, 1983, Vol. 89, № 1, P. 33-51.
[180] Ladikov Yu.P., Vasiliev V.I., Kazachkov I.V. Lining stability in steel-smelting units// *J. Soviet automatic control*, 1980, Vol.13, P. 67-72, *Transl. from J. Avtomatika*, 1980, № 4.
[181] Merkulov V.I. *Control of fluid flow*, Novosibirsk: Nauka, 1981, 174 p. (in Russian).
[182] Drazin P.G. Stability of parallel flow in a oscillating magnetic field// *Q. J. Mech.*, 1967, Vol. 20, Pt. 2, P. 201-218.

[183] Peskin R., Raco R. Ultrasonic atomization of liquids// *J. Acoust. Soc. Am.*, 1963, Vol. 35, № 9, P. 1378-1385.
[184] Ohnesorge W. Die Bildung von Tropfen an Düsen und die Auflösung flüssiger Strahlen. *Zeitschrift für Angewandte Mathematik und Mechanik*, 1936, № 16, S. 355-358.
[185] Deryagin B.V., Churaev N.V. *Wetting films*, M.: Nauka, 1984, 159 p. (In Russian).
[186] Joseph, D.D. *Stability of Fluid Motions, I, II*, New York: Springer-Verlag, 1976, 640 p.
[187] Zhukhovitsky A.A., Shatov A.A. Interaction of surface phenomena and hydrodynamic processes// *Zhurnal physkhimii (physical chemistry)*, 1985, V.59, № 10, P. 2570-2572.
[188] Frenkel Ya.I. *Kinetic theory of liquids*, Oxford: Clarendon Press, 1946, 360 p.
[189] Dussan V E.B., Davis S.H. On the motion of a fluid-fluid interface along a solid surface// *J. Fluid Mech*, 1974, Vol. 65, P. 71-95.
[190] Kern J. *Zur Hydrodynamik der Rinnsale// Verfahrenstechnik*, 1971, № 5, S. 289-294.
[191] Melcher J. Electrohydrodynamic surface waves/ *Proc. Symp. Waves Fluid Interfaces. Madison, Wisc.*, 18-20 Oct., 1982, New-York, 1983, P. 167-200.
[192] Grinchik N. N. and Grinchik Yu. N. Fundamental Problems of the Electrodynamics of Heterogeneous Media, *Physics Research International*, Vol. 2012, 28 p.
[193] Briskman V.A. Parametric stabilization of the interface of fluids, *Dokl. Akad. Nauk (Reports Acad. Sci. USSR)*, 1976, 226,№5, P.1041-1044.
[194] Troyon F., Gruber R. Theory of the dynamic stabilization of the R.-T. instability// *Phys. Fluids*, 1971, Vol. 14, № 10, P. 2069-2073.
[195] Zienkiewicz O.C., Morgan K. *Finite Elements and Approximation*, Dover Publ., 2006, 328 p.
[196] Levich V.G. Physicochemical Hydrodynamics, Englewood Cliffs, NJ: Prentice-Hall (transl. by Scripta Technica), 1962, 700 p.
[197] Zak M. Shape instability in thin viscous films and jets// *Acta Mech*, 1985, 55, №1-2, P. 33-50.
[198] Asare H.R., Takahashi R.K., Hoffman M.A. Liquid Sheet Jet Experiments: Comparison With Linear Theory// *J. Fluids Eng.*, 1981, 103 (4), P. 595-603.
[199] Muraviev E.V., Tananaev A.V., Chudov A.V., et al. MHD-flow with free surface on plate of liquid-metal diaphragm, II Riga conf. on MHD, Part I, 1984, P. 111-114 (In Russian).
[200] Aitov T.N., Ivanov A.B., Kirilina E.M. Divertor plates with a protective film/ *Intor-related Tokamak concept innovations*, Phase IIA, third part, v. II, Vienna, 1986, E.7.19.
[201] Taktarov N.G. Filtration of magnetized fluid in deformable porous medium// *J. Eng. Physics and Thermophysics*,1985, XVIII, №1, P.49-54.
[202] Sandulyak A.V., Korkhov O.V., Dakhnenko V.L., et al. The high-speed modes of filtering at magnetic cooling of particles from low-concentrated mono- and polydisperse suspenseons of a magnetic// *J. Engineering Physics and Thermophysics (IFZh)*, 1985, Vol. XVIII, № 4, P. 598-602 (In Russian).
[203] Taktarov N.G. Convection of magnetized fluids in electroconductive porous media// *J. Magnetohydrodynamics*, 1984, № 3, P. 38-40.
[204] Gorislavets Yu.M., Kazachkov I.V., Kolesnichenko A.F. Polydispersed MHD-flow in a cylindrical vessel// *Magnetohydrodynamics*, 1986, Vol. 22, № 1, P. 73-80. Transl.: Magnit. Gidrodinamika, Jan-March 1986, №1, P. 85-92.
[205] Ganichev A.I., Domashenko A.M., Nesmelov A.G., et al. Definition of a threshold of teardrop and destruction of a free surface of cryogenic liquid at vertical vibrations of capacity. *Works of MBTU Bauman*, M.: 1985, № 442, P. 33-42.
[206] Miles John W. Parametrically excited solitary waves// *J. Fluid Mech*,1984, №148, P.451-460.
[207] Brauner Neima, Maron David Moalem, Dukler Abraham E. Modeling of wavy flow in inclined thin films in the presence of interfacial shear// *Chem. Eng. Sci.*,1985,40,№6,P.923-937.
[208] Gorbunov E.V. Thermo-capillary stability of thin liquid film// *Soviet Phys. JEPT. Pis'ma Zh. Tekh. Fiz.*, 1985, Vol. 11, № 23, P. 1456-1157.
[209] Nepomnyashchiy A.A., Simanovskiy I.B. Thermocapillary convection in two-layer systems in the presence of surfactant on boundary interface// *Izv. Akad. Nauk SSSR, Mekhan. Zhidk. Gaza*, 1986, № 2, P. 3-8 (in Russian).
[210] Goussis D., Kelly R.E. Effects of viscosity variation on the stability of film flow down heated or cooled inclined surface long-wavelength analysis// *Phys. Fluids*, 1985, Vol. 28, № 11, P. 3207-3214.
[211] Velarde M.G., Castillo J.L. Thermohydrodynamic instabilities: buoyancy-thermocapillary convection. *Nonequilibr. Coop. Phenom. Phys. and Relat. Fields*/ Proc. NATO ASI, El Escorial, Madrid, Aug., I-II, 1983, New York, London, 1984, P. 179-196.
[212] Buyevich Yu.A., Kudymov S.V. To the characteristic of poorly non-linear stationary waves in a thin liquid film// *J. of Engineering Physics and Thermo-physics. IFZh*, 1983, Vol. 47, № 3, P. 566 (In Russian).
[213] Monin A.S. Hydrodynamic instability// *Sov. Phys. Usp*, 1986, Vol. 29, P. 843–868.

[214] Anno J.N. The mechanics of fluid jets, Lexington: Mass-Toronto, 1977, XIII, 102 p.
[215] Voloshyn A.Ya., Budennyi V.F., Kolesnichenko A.F., et al. Electromagnetic control of the granulation process of liquid metals by means of specialized power supply systems// *Eng. electrodynamics*, 1985, № 3, P. 3-8.
[216] Ma Yan-Chow. Wilton ripples phenomenon with a background current// *Phys. Fluids*, 1985, Vol. 28, № 4, P. 1033-1039.
[217] Baranov G.A., Khuharev V.V. *Investigation of a free ring jet flow of liquid*, Leningrad: 1980, 57 p. (preprint/NIIEFA, №П-А-0471).
[218] Genkin A.L., Kukes V.I., Yarin L.P. On the spreading of a jet of immiscible fluids, *Problems of heat power and applied thermal physics*, Alma-Ata, 1973, № 9, P. 100-104.
[219] Goussis D., Kelly R.E. Effects of viscosity variation on the stability of film flow down heated or cooled inclined surface long-wavelength analysis// *Phys. Fluids*, 1985, Vol. 28, № 11, P. 3207-3214.
[220] Lyell M.J., Huerre P. Linear and nonlinear stability of plane stagnation flow// *J. Fluid Mech.*, 1985, 161.- P. 295-312.
[221] Hooper A.P. Long-wave instability at the interface between two viscous fluids: thin layer effects// *Phys. Fluids*, 1985, Vol. 28, № 6, P. 1613-1618.
[222] Arai Takakage, Hashimoto Hiroyuki. Disintegration of thin liquid sheet in cocurrent gas stream. Wave motion of thin liquid sheet and breakup patterns// *Trans. Jap. Soc. Mech. Eng.*, 1985, B51, № 470, P. 3336-3342.
[223] Crapper G.D., Dombrowski N. A note on the effect of forced disturbances on the stability of thin liquid sheets and on the resulting drop size// *Int. J. Multiphase Flow*, 1984, Vol. 10, № 6, P. 731- 736.
[224] Davis S. Rupture of thin liquid films. Waves Fluid Interfaces/ *Proc. Symp. Madison, Wisc.*, 18-20 Oct. 1982, New York, 1983, P. 291-302.
[225] Gallez D., Prevost M. Linear and nonlinear dynamics of free liquid films// *Physicochem. Hydrodyn.*, 1985, Vol. 6, № 5-6, P. 731-745.
[226] Azuma Tsuneo, Hoshino Tatsuro. The radial flow of a thin liquid film// *Trans. Jap. Soc. Mech. Eng.*, 1984, B50, №432, 1: Laminar-turb. Trans., P. 974-981; 2: Film thickness, 982-989; 3: 1126-1133; 4: 1136-1143; 5: 3176.
[227] Linde H., Schwartz P., Wilke H. Dissipative structures and non-linear kinetics of Marangoni instability/ *Dynamics and Instability of Fluid Interfaces*, 1979, Vol. 105, P. 75-119.
[228] Lin S.P. Film waves. Waves Fluid Interfaces/ *Proc. Symp. Madison, Wisc*. 18-20 Oct, 1982.
[229] Radev S., Tchavdarov B. Spectral analysis of the Orr-Sommerfeld equation for capillary liquid jets// *Zeit. Angew. Math. und Mech.*, 1985, Vol. 65, № 4, P. 230-232.
[230] Galitseiskiy B.M., Lozhkin A.L., Ushakov A.N. Heat transfer and hydrodynamics in porous medium under oscillations of gas flow// News acad. Sci. Belorussia, Physics-energy series, Minsk, 1986, №2, P. 101-104 (In Rus.).
[231] Kazachkov I.V., Kolesnichenko A.F., Kuchinsky V.P., et al. Vibroakustische Zerstörung von Fluessigkeitsstrahlen und Duennschichten/ *Abstr. 13$^{th}$ Int. Congress on Acoustics. Yugoslavia*, Aug 24-31, 1989.
[232] Kazachkov I.V., Kolesnichenko A.F., Gorislavets Yu.M., et al. Controlled decomposition of a liquid metal jets and films in technological and power devices, *Liquid metal MHD,* Kluwer Acad. Publ. Holland, 1989, P. 293-298.
[233] Kazachkov I.V., Emets M.Yu. Electromagnetic excitation and suppression of the waves in films/ *Proc. of Int. Conf. on Energy Trans. in MHD-flows,* Cadarache, France, 1991.